\documentclass[10pt,journal,compsoc]{IEEEtran}
\usepackage{lipsum}
\usepackage{fancyhdr}
\usepackage{xcolor}
\usepackage{graphicx}
\usepackage{subfigure}
\usepackage{caption}
\usepackage{float}
\usepackage{amsmath}
\usepackage{breqn}
\usepackage{empheq}
\usepackage{url}
\usepackage{epstopdf}
\usepackage{amsmath,mleftright}
\usepackage{xparse}
\usepackage{array}
\usepackage{booktabs}
\NewDocumentCommand{\evalat}{sO{\big}mm}{%
  \IfBooleanTF{#1}
   {\mleft. #3 \mright|_{#4}}
   {#3#2|_{#4}}%
}

\usepackage[noadjust]{cite}

\ifCLASSINFOpdf
\else
\fi
\begin{document}

\title{How the Mechanical Properties and Thickness of Glass Affect TPaD Performance}

\author{Heng~Xu,~\IEEEmembership{Member,~IEEE,}
        Michael~A.~Peshkin,~\IEEEmembership{Senior~Member, IEEE}
        and~J.~Edward~Colgate,~\IEEEmembership{Fellow,~IEEE}
\IEEEcompsocitemizethanks{\IEEEcompsocthanksitem The authors are with the department of Mechanical Engineering, Northwestern University, Evanston, IL, USA 60208-3111.
\IEEEcompsocthanksitem E-mail: hengxu@u.northwestern.edu; peshkin@northwestern.edu; colgate@northwestern.edu.}
\thanks{}}

\IEEEtitleabstractindextext{%
\begin{abstract}
One well-known class of surface haptic devices that we have called TPaDs (Tactile Pattern Displays) uses ultrasonic transverse vibrations of a touch surface to modulate fingertip friction. This paper addresses the power consumption of glass TPaDs, which is an important consideration in the context of mobile touchscreens. In particular, based on existing ultrasonic friction reduction models, we consider how the mechanical properties (density and Young's modulus) and thickness of commonly-used glass formulations affect TPaD performance, namely the relation between its friction reduction ability and its real power consumption. Experiments performed with eight types of TPaDs and an electromechanical model for the fingertip-TPaD system indicate: 1) TPaD performance decreases as glass thickness increases; 2) TPaD performance increases as the Young's modulus and density of glass decrease; 3) real power consumption of a TPaD decreases as the contact force increases. Proper applications of these results can lead to significant increases in TPaD performance.

\end{abstract}

\begin{IEEEkeywords}
Ultrasonic friction reduction, mechanical properties, vibration velocity, power consumption, TPaD.
\end{IEEEkeywords}}

\maketitle
\thispagestyle{fancy}
\IEEEdisplaynontitleabstractindextext

\IEEEpeerreviewmaketitle

\section{Introduction}
\IEEEPARstart{I}{n} the quest to develop haptic interfaces for touchscreens, a number of studies have focused on tactile friction modulation based on two main technologies: electroadhesion and ultrasonic friction reduction. Electroadhesion (also known as electrovibration) uses an electric field across the skin-surface interface to create coulombic attraction and increase friction forces (\cite{bau2010teslatouch, linjama2009sense, shultz2015surface, meyer2014dynamics}). In contrast, ultrasonic friction reduction uses intermittent contact between the fingertip and a vibrating surface to lower friction (\cite{salbu1964compressible,winfield2008variable,wiertlewski2016partial}). Either method may be used to modify the friction between the fingertip and touch surface as a function of time, finger position, or other variables, thus creating tactile effects that can be readily perceived as the finger moves across the surface.

Ultrasonic friction reduction was first reported in 1995 (\cite{watanabe1995method}) and has been studied in-depth for the past decade. Biet et al. (\cite{biet2007squeeze}) employed an array of piezoelectric actuators bonded to a metal plate to create a haptic touchpad. Winfield et al. (\cite{winfield2007t}) developed the earliest TPaD (Tactile Pattern Display), which consisted of a single piezoelectric actuator bonded to a glass disk, and integrated with a finger position sensor and computer controller. Both of these early devices, and virtually every other developed since, operated at a resonant frequency in order to excite a large-amplitude standing wave.

For mobile electronic devices, power consumption is a key consideration. Unfortunately, it remains unclear what factors affect the relation between power consumption and friction reduction ability, which we term ``TPaD performance'' in this paper. High TPaD performance is expected to render high-quality haptic perception while consuming little energy. Although some researchers studied the effects of substrate material on friction reduction, heavy reliance on simulation as well as the lack of parameter variation in experimental work, make existing conclusions less convincing. Here, we design eight types of TPaDs and conduct experiments that focus on the influence of the mechanical properties and thickness of the vibrating glass on TPaD performance. In addition, we present an electromechanical model that, in combination with the experiments, elucidates the relationship between the contact force and power consumption. This approach also sheds light on the underlying mechanisms, which allows for design optimization.

\section{Experiment Platform and Method}
\subsection{Design Process}
TPaDs were made of a piece of rectangular glass (S.I. Howard Glass Company, Worcester, MA, USA) and a hard piezoelectric actuator (SMPL60W5T03R112, Steminc and Martins Inc, Miami, FL, USA). An epoxy adhesive (Acrylic Adhesive 3526 Light Cure, Loctite, Westlake, OH, USA) was used to bond the glass and the piezoelectric actuator together. For TPaDs in the experiments, energy loss can occur in various ways, including the following (\cite{wiertlewski2015power}): 1) dissipation via the artificial finger, 2) acoustic radiation into the surrounding air, 3) inelastic deformation of the TPaD, 4) dissipation via the mounting, 5) dissipation via the epoxy adhesive, 6) electric losses in the piezoelectric actuator. In this paper, we focus on the effects of glass mechanical properties and thickness on TPaD performance. Thus, other factors affecting TPaD performance were designed to be the same, including the amount of epoxy adhesive and the location of the piezoelectric actuator on the glass plate.

\begin{figure}[htb!]
\centering
\includegraphics[width = 0.4\textwidth]{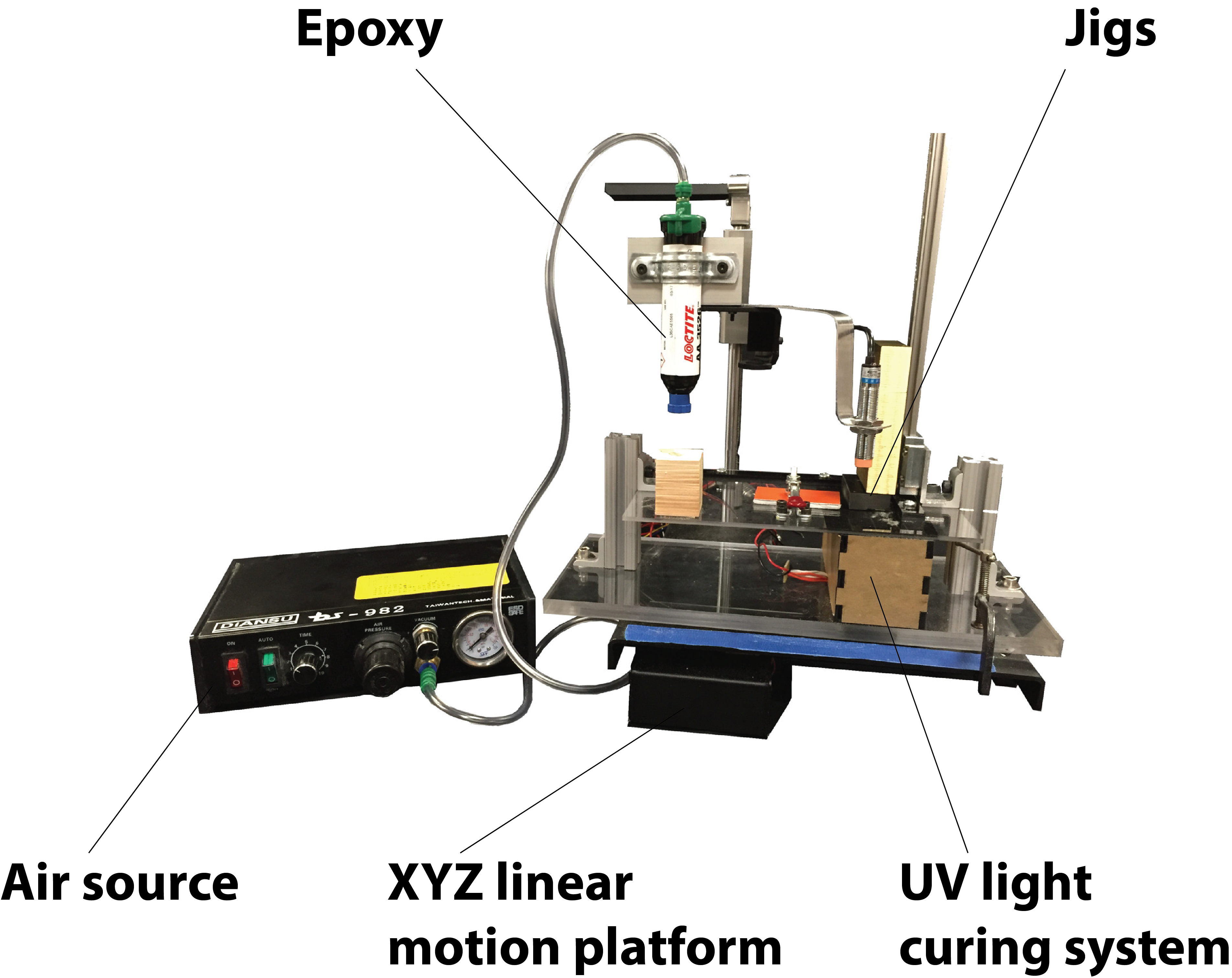}
\caption{Automatic epoxy adhesive dispenser system.}
\label{glue_dispenser.png}
\end{figure}

\begin{figure}[htb!]
\centering
\includegraphics[width = 0.4\textwidth]{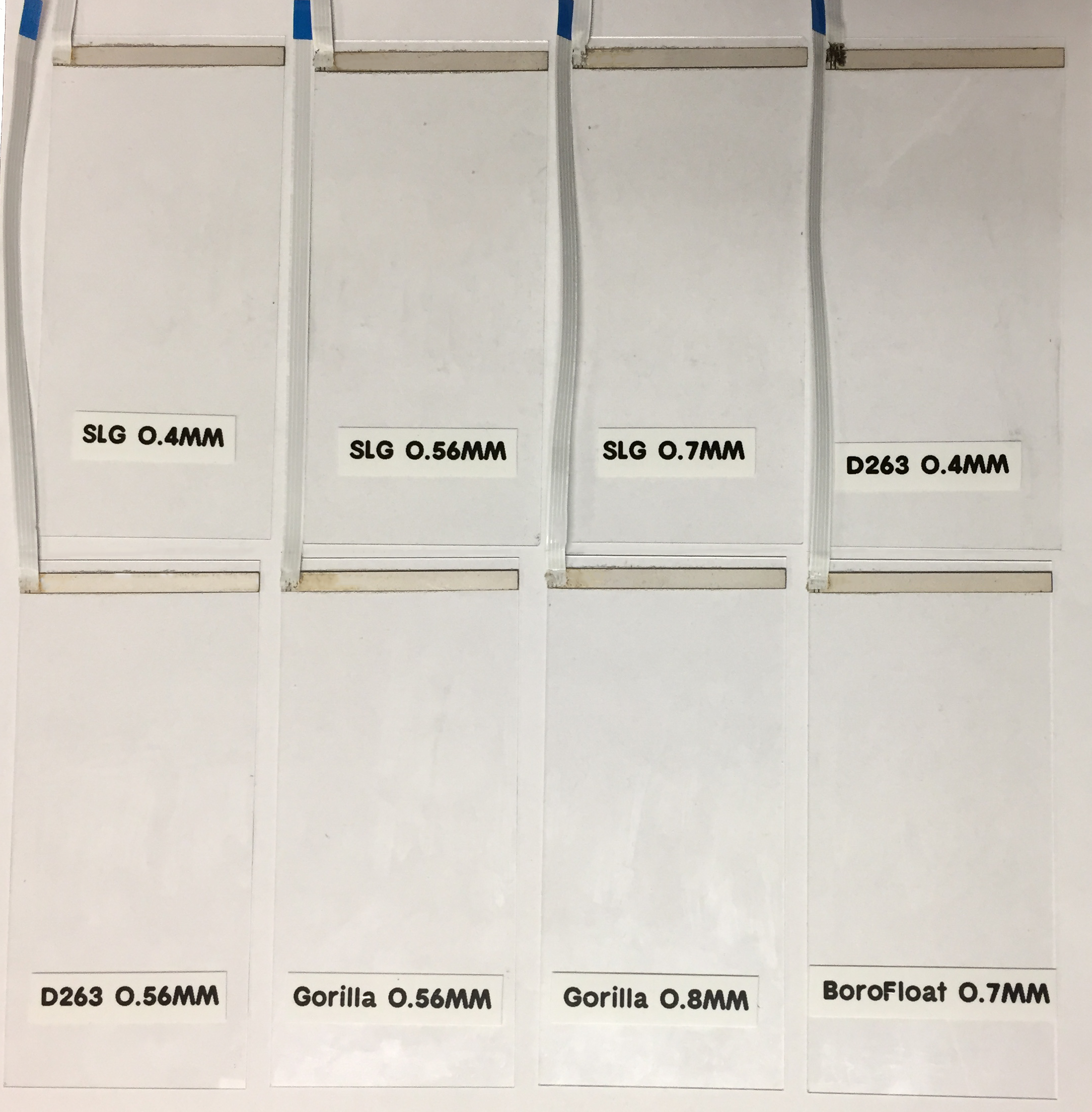}
\caption{Eight types of glass TPaDs.}
\label{TPaD_samples.png}
\end{figure}

During the design process, the shapes of the glass were identical (60 x 130 $m{m^2}$) and we changed only thickness. An automatic epoxy adhesive dispenser and jigs (shown in Fig. \ref{glue_dispenser.png}) were designed to ensure that the epoxy adhesive was dispensed identically for each assembly and that the relative position of the piezoelectric actuator on the glass was also identical across assemblies (shown in Fig. \ref{TPaD_samples.png}). The TPaDs were designed to operate in the 20 x 0 resonant mode (20 nodal lines along the length direction, but no nodal lines along the width direction).

In the following experiments, we used eight types of glass that varied in material and thickness (${l_T}$). There were four types of glass materials, including soda-lime glass, borofloat glass, D263 glass, and Gorilla glass. 

Soda-lime glass (SLG), which is composed of $Si{O_2}$, sodium oxide (soda) and calcium oxide (lime), is prepared by melting sodium carbonate, lime, dolomite, silicon dioxide, and aluminum oxide. It is usually produced for windowpanes and glass containers. Both the borofloat glass and the D263 glass belong to the family of borosilicate glasses, but they are fabricated via different processes. Borofloat glass is produced by the microfloat process, in which the top surface is fire-polished and the bottom surface floats on molten tin. In contrast, both surfaces of the D263 glass are fire-polished by a special down-draw production process. Borofloat glass is used for optoelectronics, photonics, and analytical equipment. D263 is used as cover glass in medical microscopy. Gorilla glass, which is an aluminosilicate thin sheet glass, is designed as a durable transparent substrate with high strength. It is ideal as a protective cover lens for electronic displays in cellular phones and computer screens. The density ($\rho$) and Young's modulus ($E$) of these glasses are shown in Table \ref{Eight types of glass}.

\begin{table}[htp!]
\centering
\caption{Mechanical properties of glass.}
\label{Eight types of glass}
      \begin{tabular}{cccc}
\hline
\hline

Glass &             Thickness   & Density & Young's modulus   \\

           &              (mm)           &    ($g/c{m^3}$) &   $kN/m{m^2}$                \\
\hline
 \# 1(SLG\_0.4) &           0.4 &           2.483 & 71 \\

 \# 2(SLG\_0.56) &           0.56 &              2.483 & 71\\

 \# 3(SLG\_0.7) &          0.7 &          2.483 & 71\\

 \# 4(D263\_0.4) &           0.4 &            2.51 & 72.9 \\

 \# 5(D263\_0.56) &           0.56 &           2.51 & 72.9 \\

 \# 6(Gorilla\_0.56) &           0.56 &          2.42 & 71.5\\

 \# 7(Gorilla\_0.8) &           0.8 &          2.42 & 71.5\\

 \# 8(BoroFloat\_0.7) &           0.7 &          2.2 & 64\\

\hline
\hline
\end{tabular}
\end{table}

\subsection{Experiment Method}
In order to ensure a consistent modal shape across samples, the excitation frequencies of the TPaDs varied from 22.4 kHz to 44.6 kHz, due to the different mechanical properties (density and Young’s modulus) and thickness of glass. These resonant frequencies were found without an artificial finger in contact. The relation between the ultrasonic friction reduction ability and the corresponding real power consumption was used to evaluate TPaD performance. Because friction reduction is variable across people and because no widely-accepted artificial finger design exists, we chose to measure the amplitude and frequency of the TPaDs and then to relate these measurements to friction reduction based on the model developed by Vezzoli, et al. and Sednaoui, et al. This approach will be further discussed in Section \ref{friction_model_review}.

\subsubsection{Experiment 1 with an Artificial Finger}
\begin{figure}[htb!]
\centering
\includegraphics[width = 0.4\textwidth]{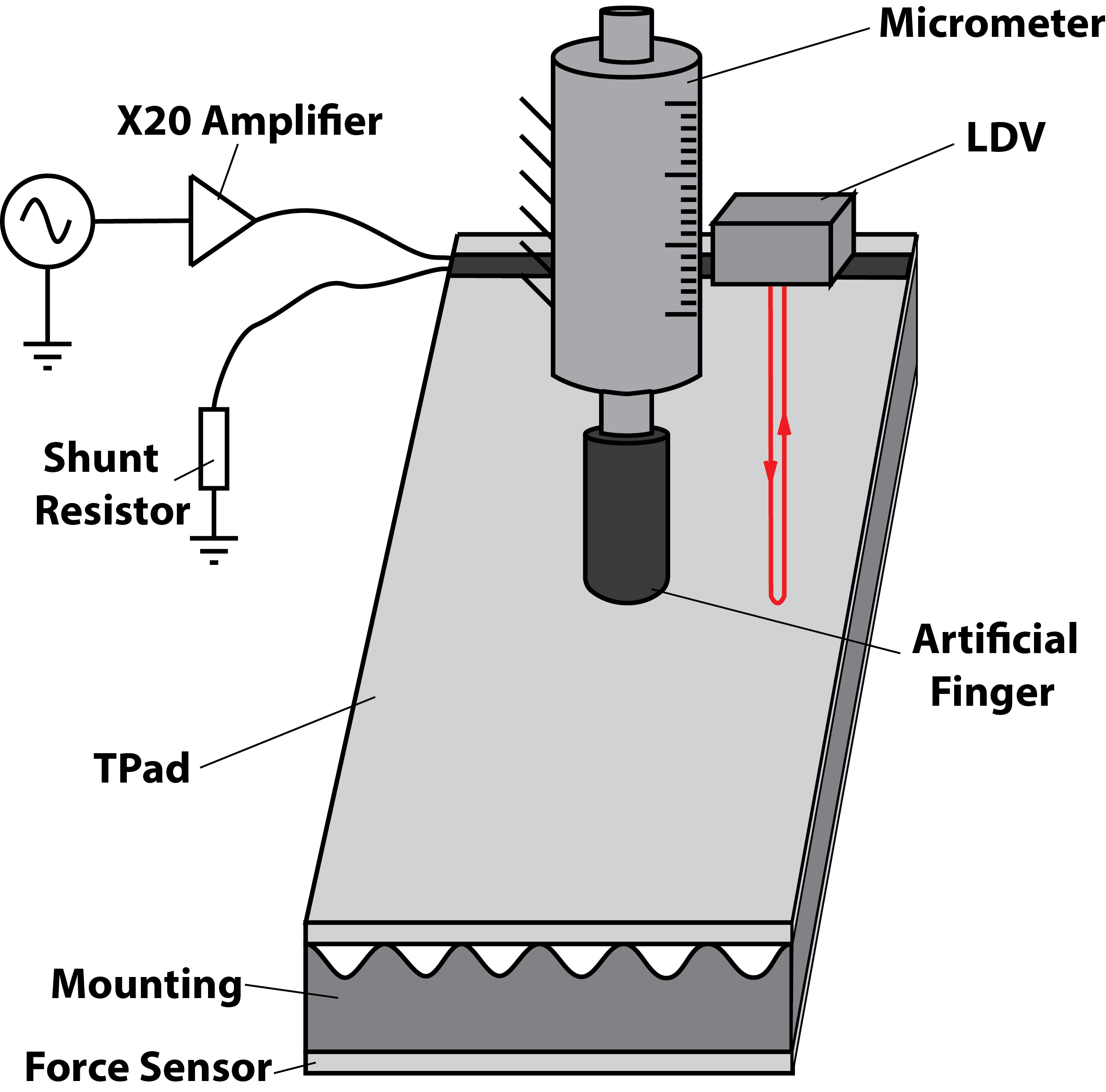}
\caption{Experiment platform.}
\label{ExperimentPlatform_Finger.png}
\end{figure}

The setup in Experiment 1 is shown in Fig. \ref{ExperimentPlatform_Finger.png}. TPaDs were placed on a piece of foam with a free boundary condition. A voltage source ($\pm$40 v) was applied to the TPaDs through a 20 times amplifier. A shunt resistor ($100\Omega \pm 0.1\%$), which was in series with the piezoelectric actuator, was used to measure the input current to the piezoelectric actuator, enabling a calculation of real power consumption. In order to measure TPaD performance under different contact forces, an artificial finger (the ``TangoPlus'' used in \cite{friesen2016role}) was used to press the glass. A micrometer and the force sensor were used to control the contact force precisely. The range of contact force was from 0 to 2 N with nine levels (0, 0.2, 0.4, 0.5, 0.6, 0.8, 1.0, 1.5, and 2.0 N). A contact force of 0 N means that there was no contact between the artificial finger and the TPaD. A laser Doppler vibrometer (LDV) was used to measure the vibration amplitude of the TPaD. The contact point between the TangoPlus and the glass was at the same antinodal line as the measurement point of the LDV. Additionally, a dynamic signal analyzer (HP35665A, Hewlett-Packard software company, Washington, USA) was used to measure the electrical impedance of the TPaD, which was to be used in the fingertip-TPaD model analysis (in Section \ref{finger_tpad_model}).

Each contact force was repeated five times. Thus, there were 45 trials for each TPaD in total. In each trial, the voltage and current applied to the piezoelectric actuator and the vibration amplitude of the TPaD were recorded for 0.1 seconds using a NI USB-6361 with a 300 kHz sampling frequency.

\subsubsection{Experiment 2 with a Spring}
The setup in Experiment 2 was similar to that in Experiment 1, except for using a spring to contact the TPaDs in order to mimic a finger without viscosity. The stiffness of the spring was $360 \pm 12 N/m$. The contact between the spring and the glass was a circle, and an epoxy adhesive was used to bond them. Based on this setup, we repeated the experiments.

\section{Ultrasonic friction reduction model review} \label{friction_model_review}
Even though researchers have broad agreement that ultrasonic friction reduction stems from intermittent contact \cite{wiertlewski2016partial,vezzoli2017friction}, it is still unclear precisely how the effect depends on the amplitude, velocity, and acceleration of the transverse vibrations. Three leading hypotheses are discussed in this section.

Based on Boyle's law and multiscale contact theory \cite{persson2013contact}, Wiertlewski, et al. \cite{wiertlewski2016partial} concluded that the relative friction force ($\mu '$), which is a ratio of the friction force when the ultrasonic vibration is on to the friction force when the ultrasonic vibration is off, depends on the square of vibration amplitude (Eq. \ref{eq:1_1}).
\begin{equation}
\begin{array}{l}
\displaystyle \mu ' \propto \exp ( - \frac{{5{\alpha ^2}{p_0}}}{{4u_0^2{p_s}}})
\end{array}
\label{eq:1_1}
\end{equation}
where the $\alpha$, $p_0$, $u_0$, and $p_s$ are the vibration amplitude of the plate, the atmospheric pressure, the gap at rest, and the pressing pressure, respectively. Although the model showed good fit with the results from human fingers and an artificial finger, it did not consider the effect of vibration frequency, which was shown to be an important factor in \cite{vezzoli2017friction,sednaoui2017friction,giraud2018evaluation}.

Vezzoli, et al. and Sednaoui, et al. \cite{vezzoli2017friction,sednaoui2017friction} employed a finite element model and spring-slider model to study effects of four factors on friction reduction: finger exploration velocity, contact force, vibration frequency and amplitude of the plate. They argued that the relative friction force depended on the velocity of the vibrating plate (Eqs. \ref{eq:1_2} and \ref{eq:1_3}). 
\begin{equation}
\begin{array}{l}
\displaystyle \mu ' = 1 - \exp ( - \frac{\Psi }{{{\Psi ^*}}})
\end{array}
\label{eq:1_2}
\end{equation}

\begin{equation}
\begin{array}{l}
\displaystyle \Psi = \frac{U}{{f\alpha {\mu _0}(1 + \nu )}}
\end{array}
\label{eq:1_3}
\end{equation}
where $U$, $f$, $\alpha$, $\mu _0$, $\nu$, $\Psi$, and $\Psi ^*$ are finger exploration velocity, vibration frequency, vibration amplitude, friction coefficient when the ultrasonic vibration is off, Poisson's ratio of the finger skin, a dimensionless group, and the characteristic value of $\Psi$.

Additionally, Giraud, et al. \cite{giraud2018evaluation} investigated the friction reduction on a vibrating plate at 66 kHz and 225 kHz. Both the friction measurements and psychophysical results showed that the ultrasonic friction reduction depended on the plate's acceleration. 

Although these three studies appear to be in conflict, closer examination suggest that all three may be reasonable within specific parameter ranges. In the experiments from Wiertlewski, et al. \cite{wiertlewski2016partial}, the vibration frequency was around 29 kHz, and the vibration amplitudes were from 0 to 3 $\mu m$. The model of Vezzoli, et al. and Sednaoui, et al. \cite{vezzoli2017friction,sednaoui2017friction} showed good agreement with experimental results in which the vibration frequencies were 10 kHz to 100 kHz and the vibration amplitudes were larger than 1 $\mu m$. The experiments of Giraud, et al. \cite{giraud2018evaluation} compared only two vibration frequencies (66 kHz and 225 kHz).

\begin{figure}[htb!]
\centering
\includegraphics[width = 0.45\textwidth]{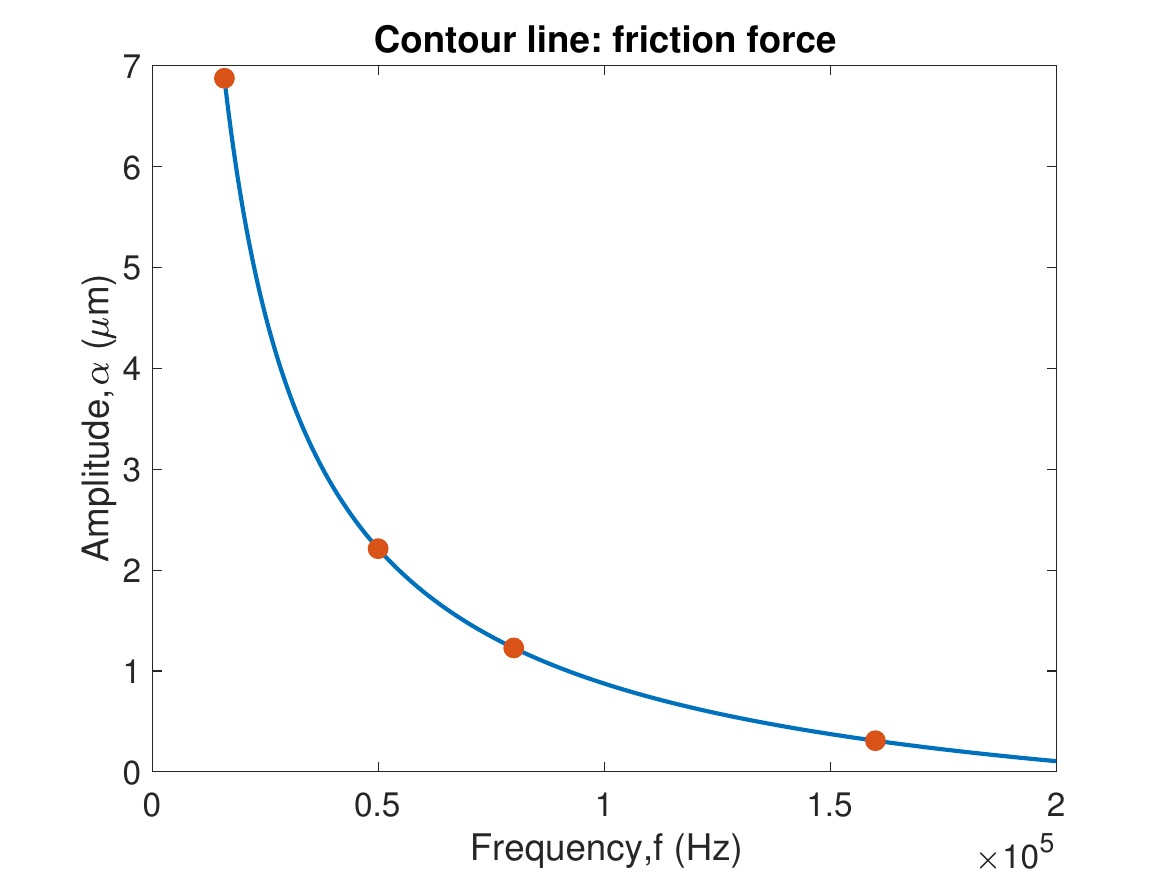}
\caption{Hypothetical friction force contour based on the vibration frequency and amplitude. The blue curve is one friction force contour, in which $\alpha = 1.755 \times 10^{4} * {f^{ - 0.797}} - 0.937$. Four red points are (16 kHz, 6.87 $\mu m$), (50 kHz, 2.21 $\mu m$), (80 kHz, 1.23 $\mu m$), and (160 kHz, 0.31 $\mu m$), respectively.}
\label{hypothesis_frequencyAmplitude}
\end{figure}

To illustrate that all of these results may be consistent with a single underlying mechanism, consider the hypothetical friction force contour shown in Fig. \ref{hypothesis_frequencyAmplitude}. In the range of 16 kHz to 50 kHz, the friction force depends largely on the vibration velocity ($f*\alpha$), while in the range of 80 kHz to 160 kHz, the friction force is close to a function of the vibration acceleration ($f^2*\alpha$). Thus, a nonlinear dependence on frequency may explain the conflict between the model of \cite{vezzoli2017friction} and the results of \cite{giraud2018evaluation}. While the nonlinear relation between vibration frequency and amplitude in Fig. \ref{hypothesis_frequencyAmplitude} is purely hypothetical, something similar could arise due to the differing effects of finger mass and viscoelastic behavior on mechanical impedance as vibration frequency and amplitude change. 

In our Experiment 1, the vibration frequencies were from 22.4 kHz to 44.6 kHz, and the vibration amplitudes without the artificial finger contact were from 1.8 to 6 $\mu m$. Since these experimental parameters fall into the range that is captured well by the model of Vezzoli, et al. and Sednaoui, et al. \cite{vezzoli2017friction,sednaoui2017friction}, Eqs. \ref{eq:1_2} and \ref{eq:1_3} will be used here to describe the relation between the vibration velocity and the relative friction force. Additionally, we note that our previous work \cite{xu2019ultrashiver} also showed that the damping of the human finger dominated the mechanical impedance in the shear direction when the finger contacted a laterally vibrating plate with 5 $\mu m$ amplitude at 30 kHz. As a result, in that work the velocity of the oscillation had a greater impact on the reaction force applied to the fingertip than either the amplitude or acceleration of the oscillation.

\section{Results}
\begin{figure}[htb!]
  \centering
  \subfigure[]{ 
    \label{all:subfig:a} 
    \includegraphics[width=0.45\textwidth]{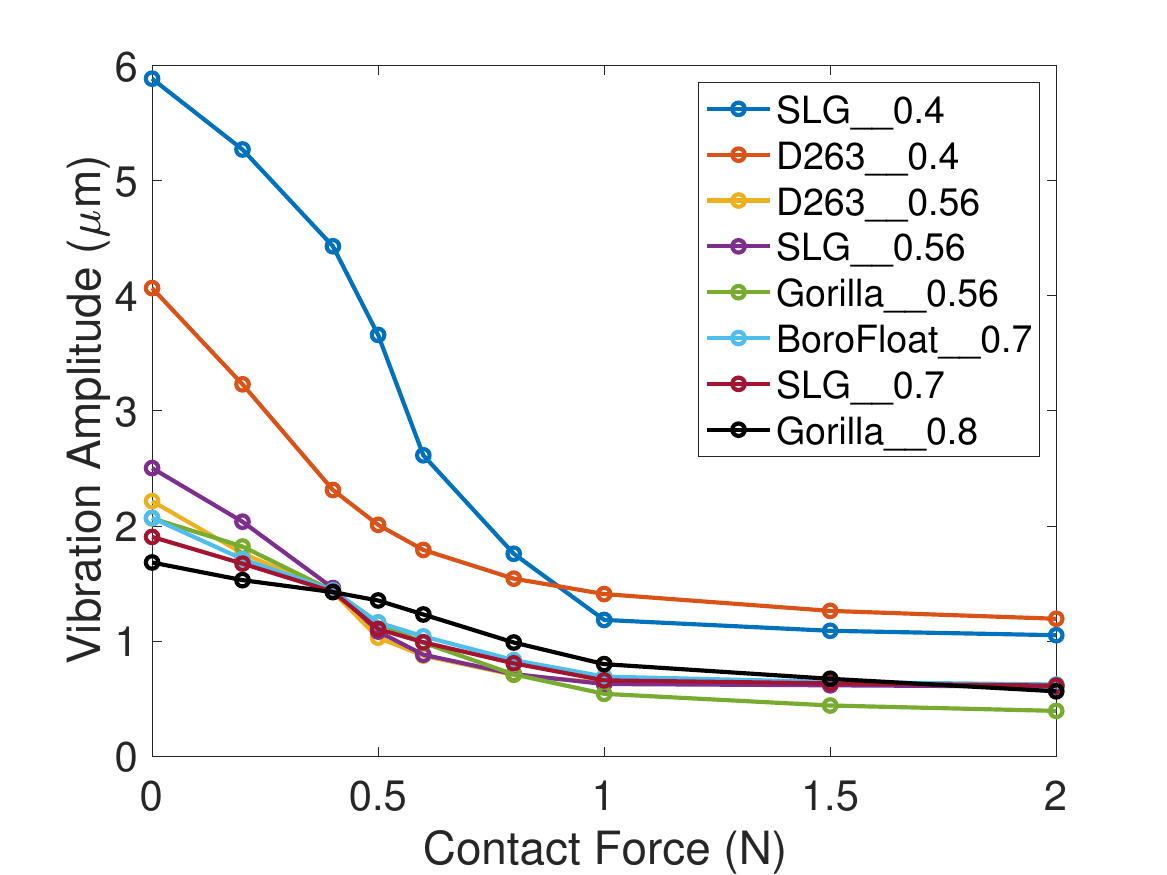}} 
  \hspace{0in} 
  \subfigure[]{ 
    \label{all:subfig:b} 
    \includegraphics[width=0.45\textwidth]{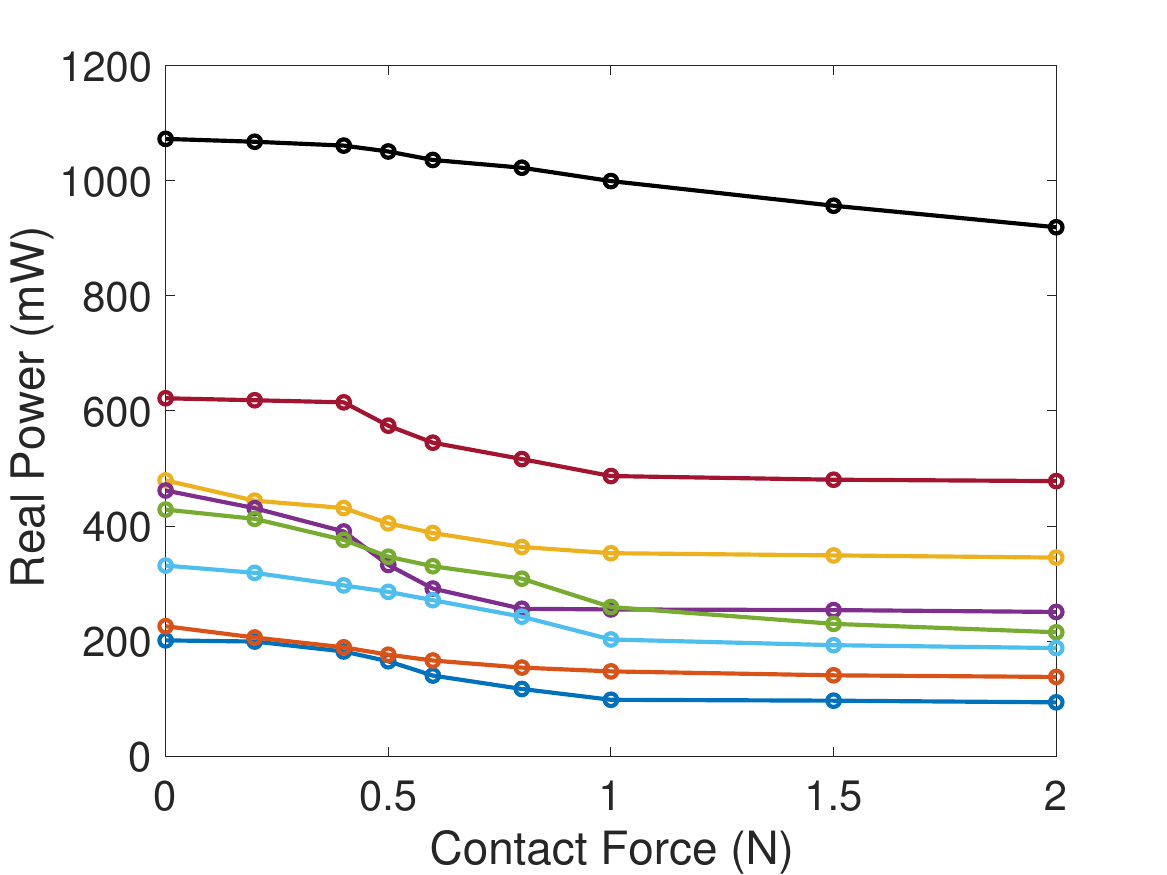}} 
  \caption{Comparisons of eight TPaDs as a function of contact force: (a) vibration amplitude, and (b) real power.}
  \label{all:subfig} 
\end{figure}

Since the data from the TangoPlus was remarkably consistent among the trials, the average values over the five trials are shown in the Figs. \ref{all:subfig}-\ref{physical:subfig} without error bars. The relative friction force in Figs. \ref{materials:subfig} and \ref{thickness:subfig} was calculated based on Eqs. \ref{eq:1_2} and \ref{eq:1_3}, where the vibration frequency ($f$) and amplitude ($\alpha$) were measured in Experiment 1, and the rest variables ($U$ = 50 mm/s, $\mu _0$ = 0.25, $\nu$ = 0.33, $\Psi ^*$ = 4.69) were taken from \cite{vezzoli2017friction,sednaoui2017friction}.

Without the contact of the artificial finger, the vibration amplitudes of TPaDs vary from 1.8 to 6 $\mu m$ (in Fig. \ref{all:subfig:a}). The SLG\_0.4 has the highest vibration amplitude (about 6 $\mu m$), which is about 2 $\mu m$ more than the second-highest vibration amplitude (D263\_0.4). As the contact force increases to 1 N, the vibration amplitudes of the TPaDs decrease significantly, especially for the SLG\_0.4, thus also decreasing the friction reduction performance (Figs. \ref{materials:subfig} and \ref{thickness:subfig}).

Compared to the sharp changes of the vibration amplitude, the real power consumption decreases slightly when increasing the contact force (in Fig. \ref{all:subfig:b}). The Gorilla\_0.8 has the highest real power consumption (around 1 Watt). The SLG\_0.4 has the lowest real power consumption, less than 0.2 Watt. Also, we observe that the power consumption of a TPaD decreases as the contact force increases (in Fig. \ref{all:subfig:b}). This will be discussed in more detail in Section \ref{Counterintuitive_section}.

\subsection{Effect of Thickness}
\begin{figure}[htbp!]
  \centering
  \subfigure[]{ 
    \label{materials:subfig:a} 
    \includegraphics[width=0.45\textwidth]{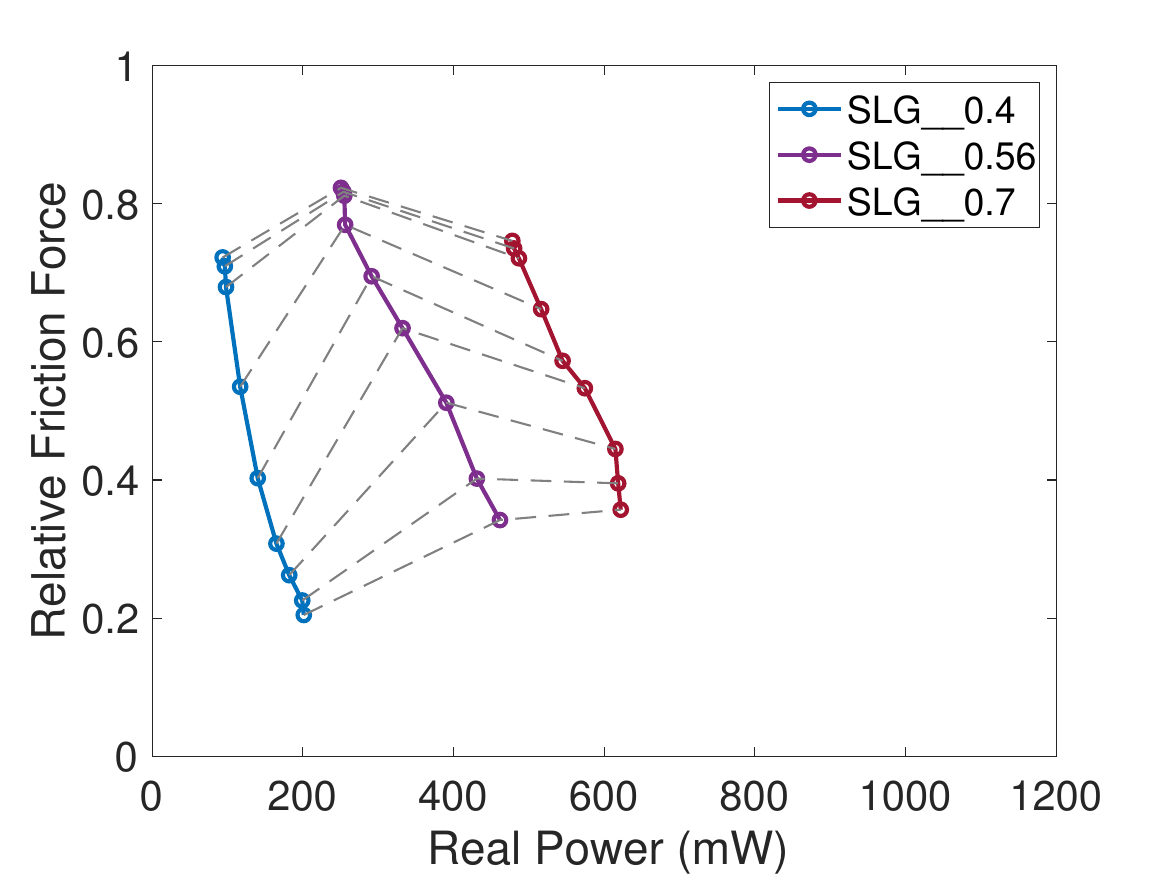}} 
  \hspace{0in} 
  \subfigure[]{ 
    \label{materials:subfig:b} 
    \includegraphics[width=0.45\textwidth]{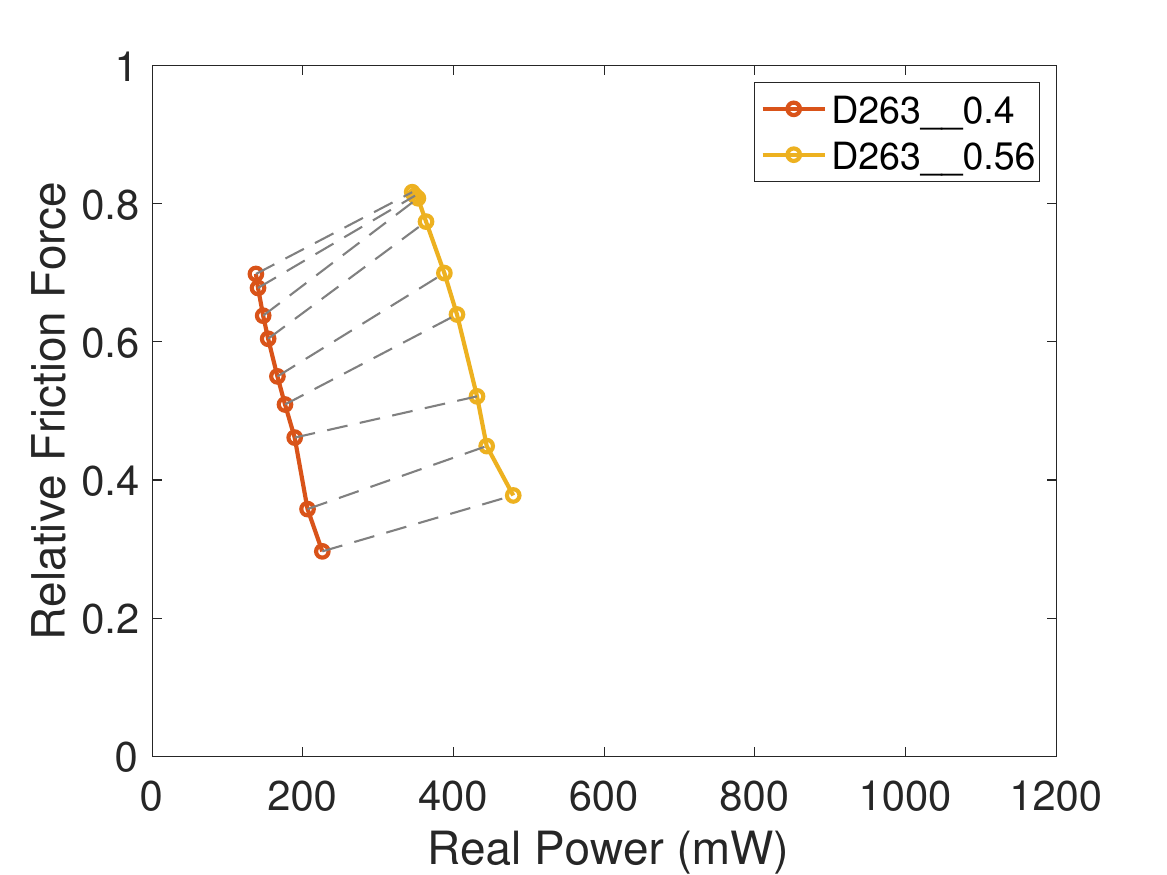}} 
  \hspace{0in} 
  \subfigure[]{ 
    \label{materials:subfig:c} 
    \includegraphics[width=0.45\textwidth]{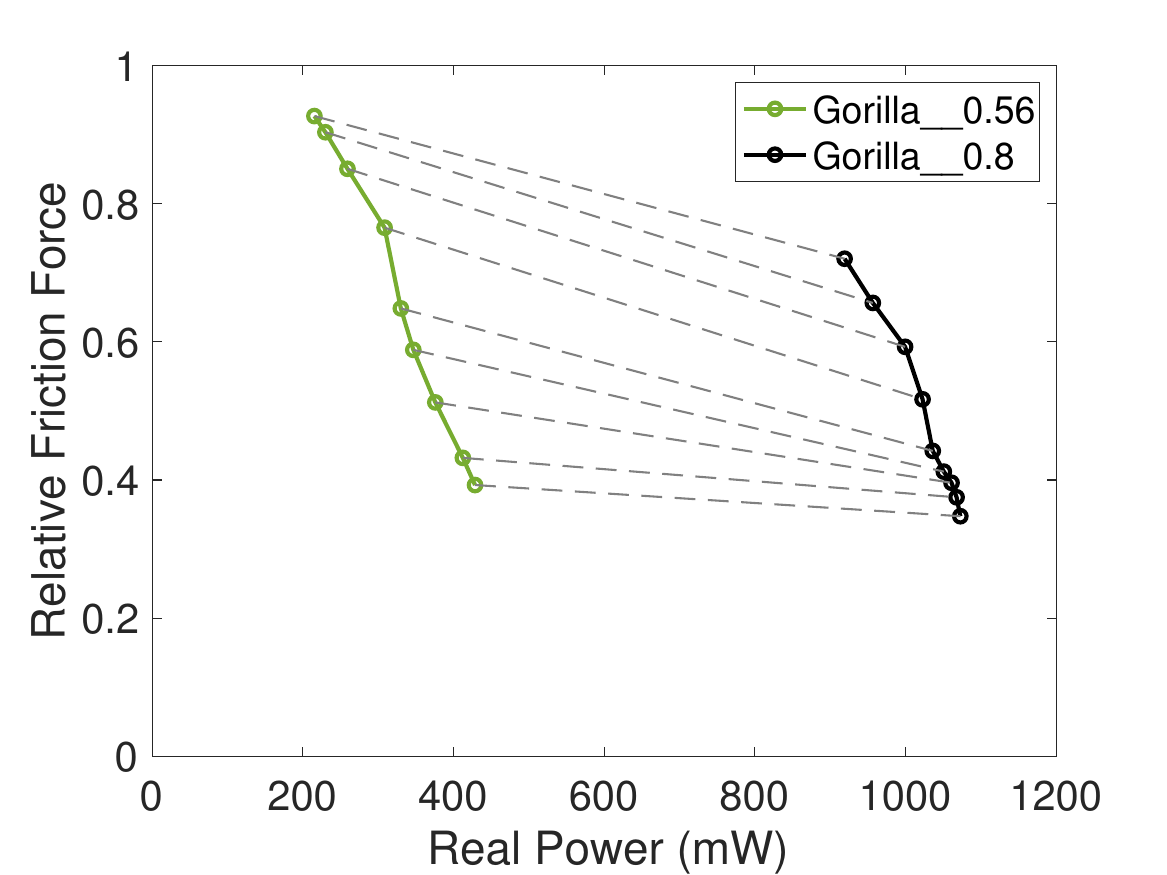}} 
  \caption{Comparisons between TPaDs with different thicknesses on the relationship between the relative friction force and the real power consumption: (a) SLG TPaDs, (b) D263 TPaDs, and (c) Gorilla TPaDs. Each point represents an average value over five trials for the corresponding TPaD. Each dashed line connects the results with the same contact force. Points with higher contact force are up and to the left.}
  \label{materials:subfig} 
\end{figure}

\begin{figure}[htbp!]
  \centering
  \subfigure[]{ 
    \label{thickness:subfig:a} 
    \includegraphics[width=0.45\textwidth]{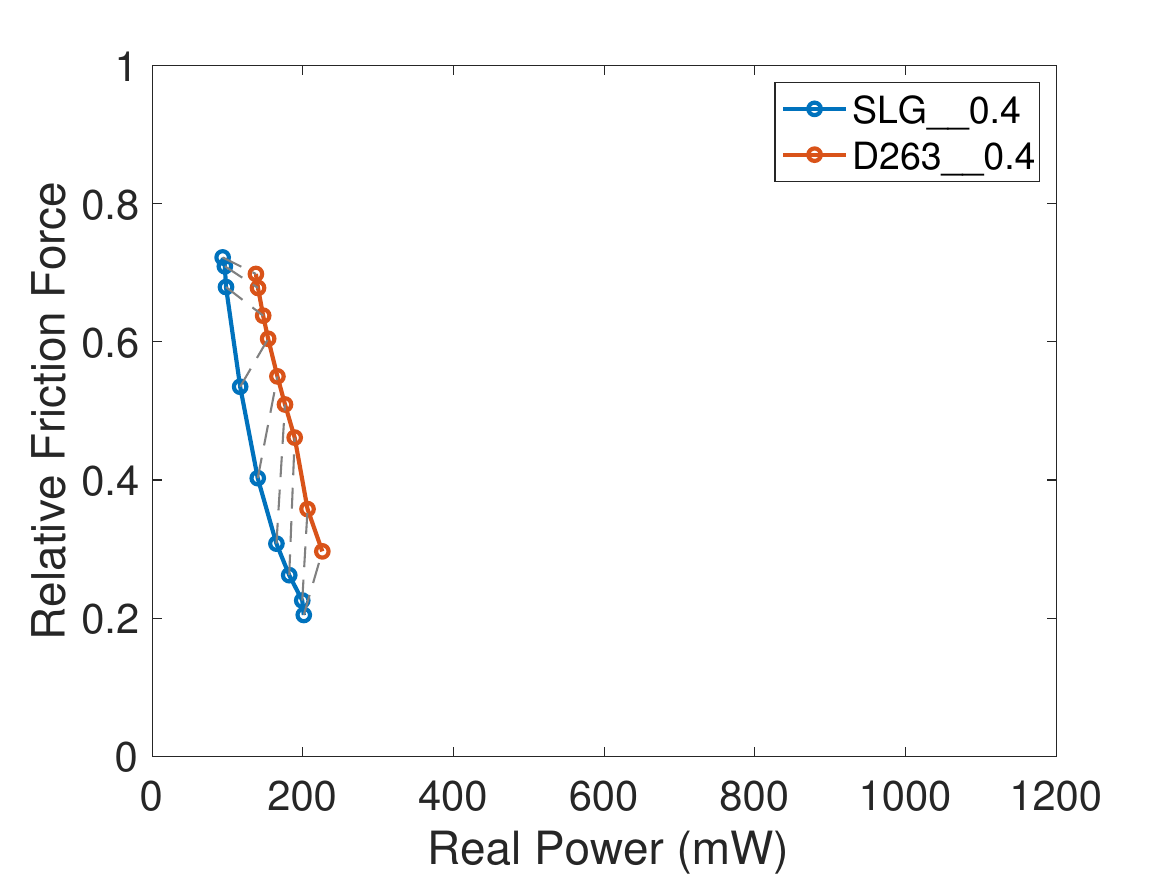}} 
  \hspace{0in} 
  \subfigure[]{ 
    \label{thickness:subfig:b} 
    \includegraphics[width=0.45\textwidth]{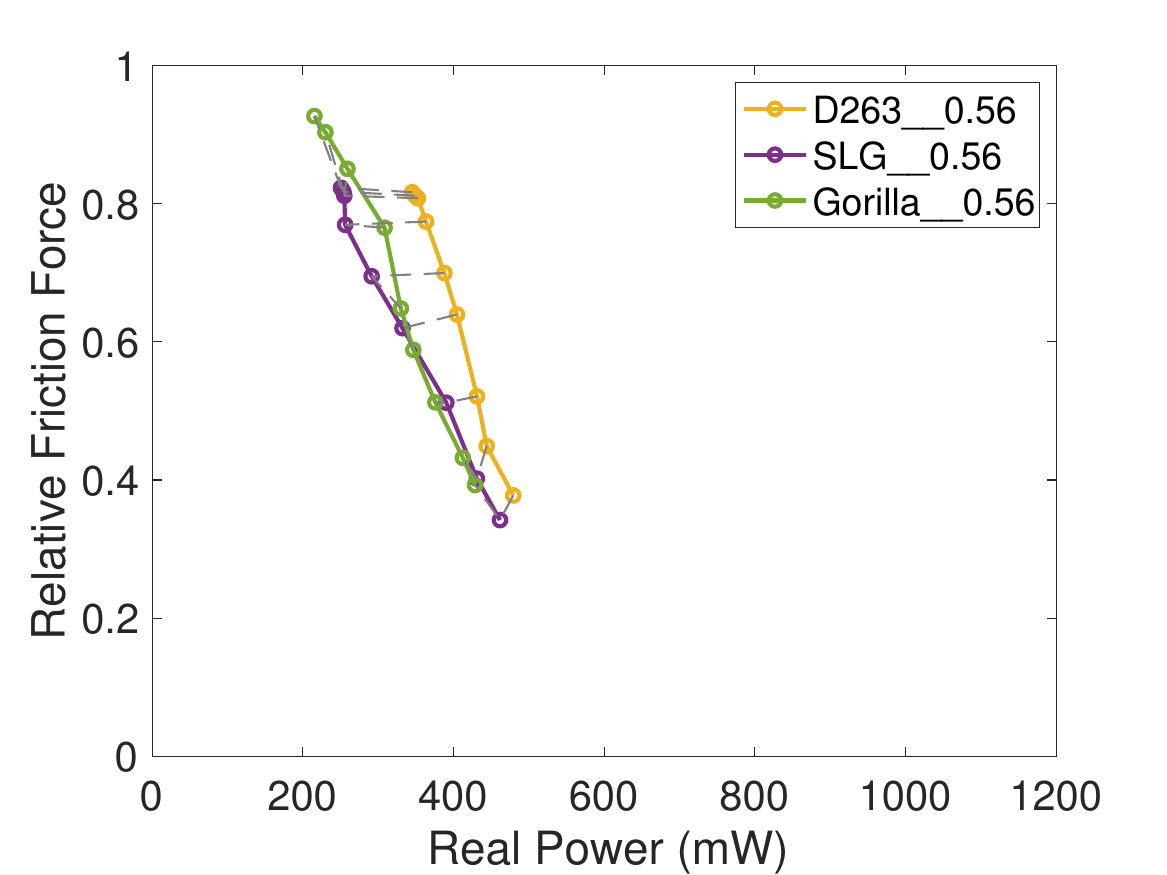}} 
  \hspace{0in} 
  \subfigure[]{ 
    \label{thickness:subfig:c} 
    \includegraphics[width=0.45\textwidth]{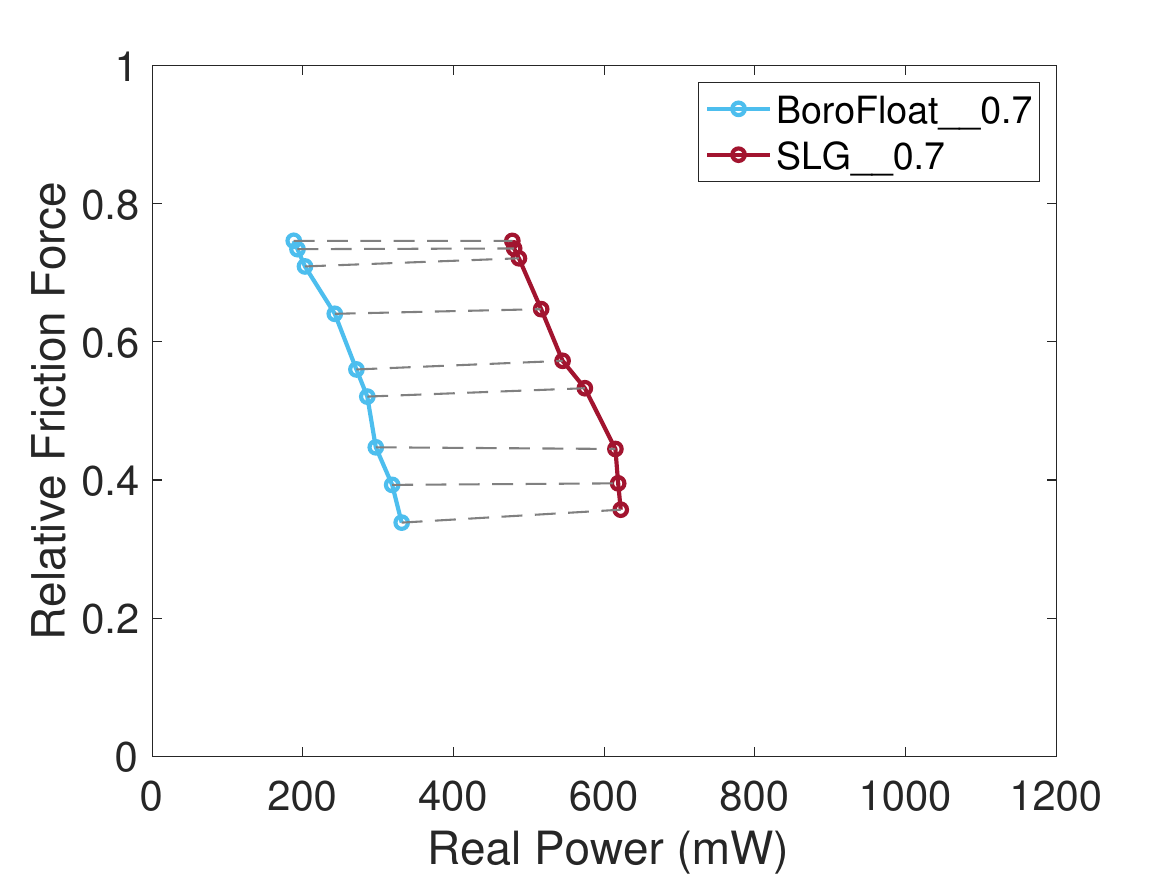}} 
  \caption{Comparisons between TPaDs with different materials on the relationship between the relative friction force and the real power consumption: (a) TPaDs with 0.4 mm thickness, (b) TPaDs with 0.56 mm thickness, and (c) TPaDs with 0.7 mm thickness. Each point represents an average value over five trials for the corresponding TPaD. Each dashed line connects the results with the same contact force. Points with higher contact force are up and to the left.}
  \label{thickness:subfig} 
\end{figure}

Fig. \ref{materials:subfig} shows the effect of thickness for each of the three materials (SLG, D263, and Gorilla). Independent of material, thinner TPaDs yield higher performance. For example, in Fig. \ref{materials:subfig:a}, SLG\_0.4 generates the maximum friction reduction with the minimum real power consumption for each contact force. SLG\_0.7 has higher friction reduction ability than SLG\_0.56, but this comes at a higher energy cost. Similar trends are also found in Fig. \ref{materials:subfig:b}. Finally, as seen in Fig. \ref{materials:subfig:c}, even though the friction reduction ability of the Gorilla\_0.8 is slightly better than that of the Gorilla\_0.56, this again comes at a significant cost in energy consumption.

\subsection{Effect of Materials}
Fig. \ref{thickness:subfig} shows the effect of material for each of the three thicknesses (0.4 mm, 0.56 mm and 0.7 mm). Here, the results are a bit more mixed. Fig. \ref{thickness:subfig:a} shows that SLG\_0.4 has lower energy consumption and comparable or better friction reduction than D263\_0.4 at each contact force. This trend holds true at 0.56 mm (Fig. \ref{thickness:subfig:b}) with Gorilla\_0.56 being similar to, but exhibiting slightly less friction reduction, than soda lime. Fig. \ref{thickness:subfig:c} shows that BoroFloat\_0.7 has almost identical friction reduction ability to SLG\_0.7 but that it costs considerably more energy.

\section{Relation Between Power Consumption and Contact Force} \label{Counterintuitive_section}
Experiment 1 shows that TPaD power consumption decreases as contact force increases. One possible reason is that the resonant frequency of a TPaD shifts as the contact force grows. Since the fixed excitation frequency was used for each TPaD in Experiment 1, the decreasing power consumption could be due to the deviation from resonance. To address this, another set of experiments was run in which the resonant frequency was updated for each contact force, and we still found similar decreases in real power consumption with increasing contact force. To investigate matters further, we built a lumped-parameter electromechanical model, which is introduced in the next section.

\subsection{Model of Fingertip-TPaD System} \label{finger_tpad_model}
\begin{figure}
  \centering
  \subfigure[]{ 
    \label{fig:subfig:a} 
    \includegraphics[width=0.3\textwidth]{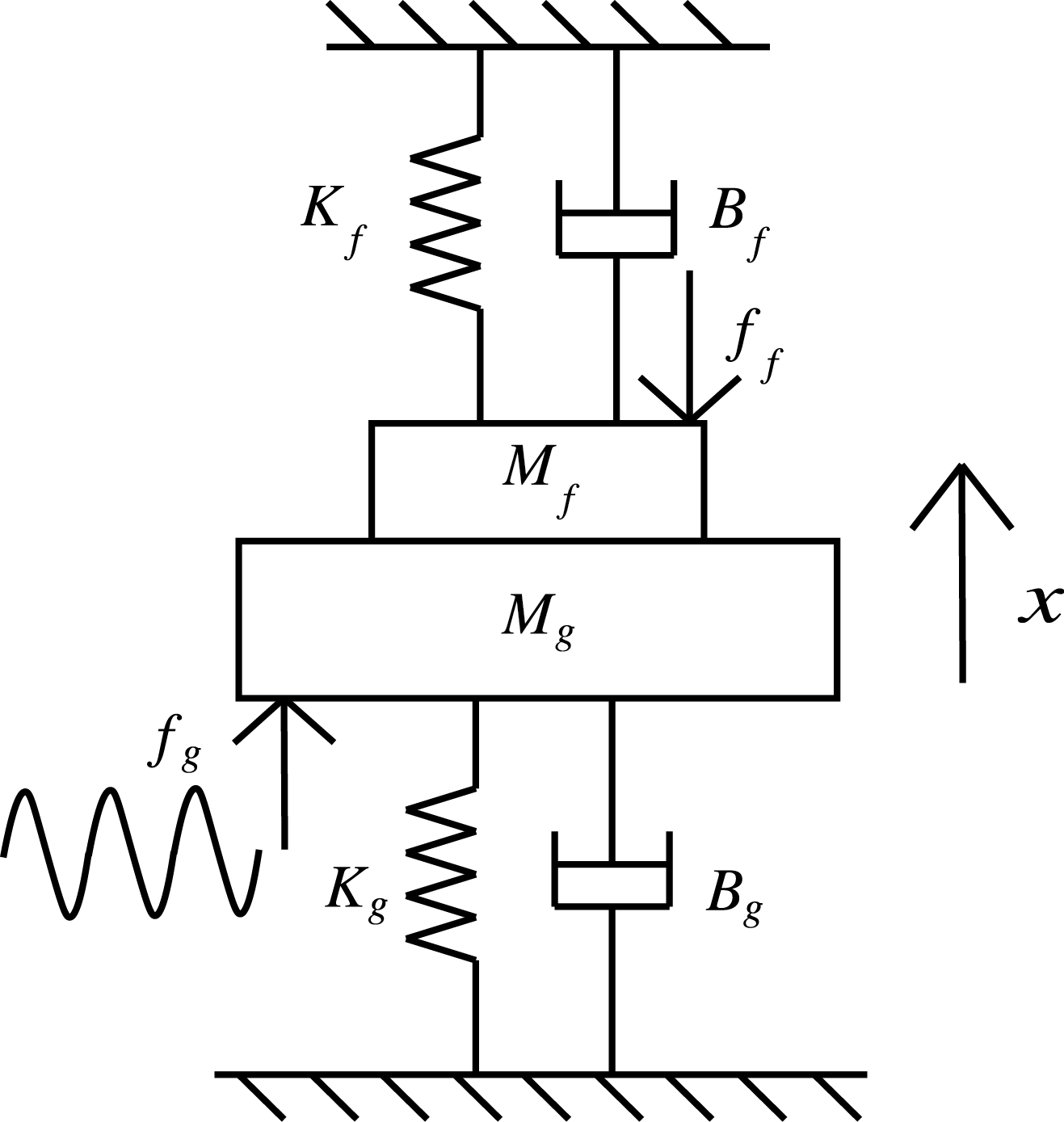}} 
  \hspace{0in} 
  \subfigure[]{ 
    \label{fig:subfig:b} 
    \includegraphics[width=0.45\textwidth]{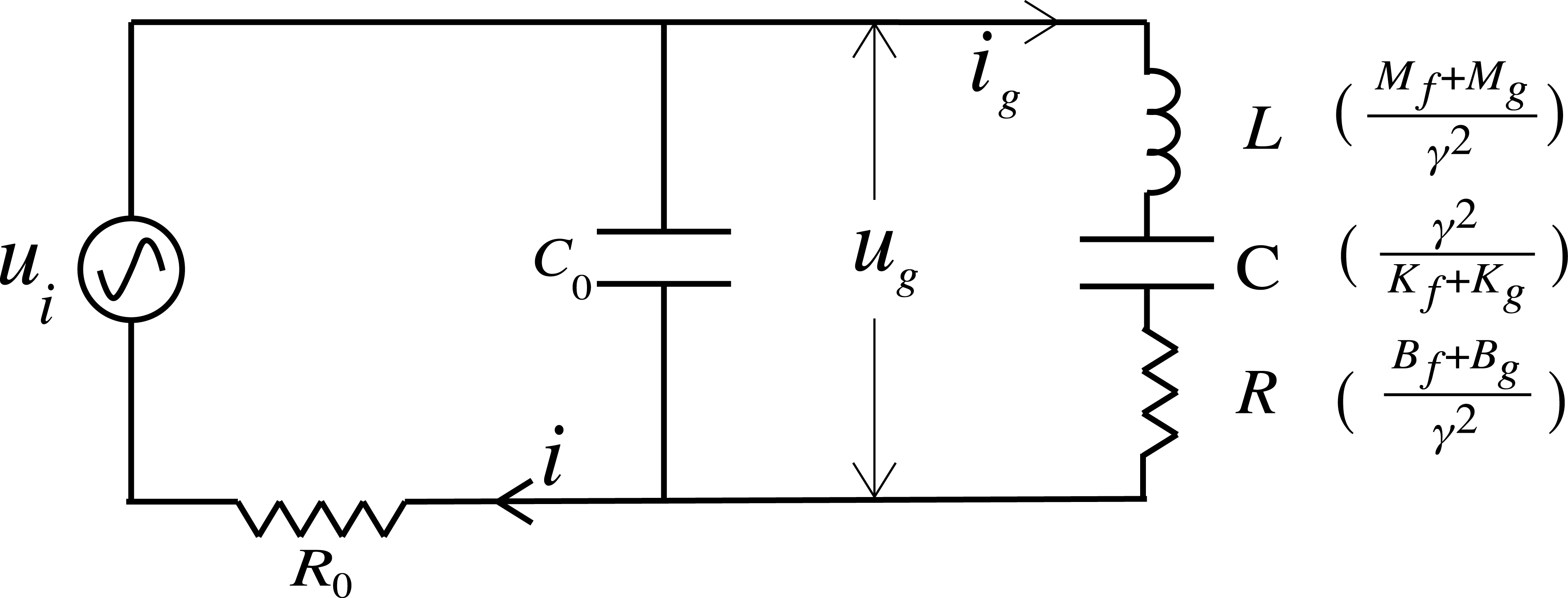}} 
  \caption{Electromechanical model: (a) lumped-parameter model and (b) equivalent electrical model. In Fig. \ref{fig:subfig:a}, the artificial finger is represented by the mass (${M_f}$), the spring (${K_f}$), and the damper (${B_f}$). The TPaD with the mounting foam is represented by the mass (${M_g}$), the spring (${K_g}$), and the damper (${B_g}$). ${f_f}$ represents the contact force between the artificial finger and the TPaD. ${f_g}$ represents the driving force, which is transformed from the constant voltage source. In Fig. \ref{fig:subfig:b}, ${u_i}$, ${R_0}$, and ${C_0}$ represent the constant voltage source, the shunt resistor, and the static capacitor of the piezoelectric actuator, respectively. $\gamma$ represents the electromechanical conversion factor. $L$, $C$, and $R$ represent the equivalent inductor, capacitor, and resistor from the lumped-parameter model, respectively. The vibration velocity ($\dot X$) is represented by ${i_g} = \gamma \dot X$, the current that passes through $L$, $C$, and $R$. The driving force ($f_g$) is represented by ${u_g }= f_g /\gamma$, the cumulative voltage across $L$, $C$, and $R$. ${i_g}$ and $i$ are the current through the motional branch of the TPaD and the total current, respectively.}
  \label{fig:subfig} 
\end{figure}

The simplified lumped-parameter model (Fig. \ref{fig:subfig:a}) includes the vibrating TPaD, the artificial finger, and the electrical behavior of the piezoelectric actuator. For present purposes, we assume the fingertip stays in contact with the TPaD when the vibration amplitude is low. In the equivalent circuit model (Fig. \ref{fig:subfig:b}), an equivalent inductor ($L$), capacitor ($C$), and resistor ($R$), which represent the impedance of artificial finger and the TPaD in the lumped-parameter model, are reflected on the electrical circuit of the piezoelectric actuator by the electromechanical conversion factor ($\gamma$). The static capacitance (${C_0}$) of the piezoelectric actuator is 9.88 nF and the resistance of the shunt resistor (${R_0}$) is $100\Omega \pm 0.1\%$. $u_g$ and $i_g$ are the voltage and the current across the motional branch of the TPaD, respectively. Other parameters were estimated based on a measurement of the electrical impedance of the fingertip-TPaD system. In addition, the pressing force (${f_f}$) is a static force, which is not included in the dynamic analysis of this model. But, the effect of the pressing force (${f_f}$) is reflected on the changes of the equivalent inductor ($L$), capacitor ($C$), and resistor ($R$) because a higher pressing force introduces more finger tissue to the finger-TPaD system.

The transfer function from $i(t)$ to ${u_g(t)}$ is equal to: 
\begin{equation}
\begin{array}{l}
\displaystyle \frac{{{U_g(s)}}}{I(s)} = \frac{{LC{s^2} + RCs + 1}}{{{C_0}LC{s^3} + {C_0}RC{s^2} + ({C_0} + C)s}}
\end{array}
\label{eq:2}
\end{equation}
from which impedance of the TPaD system can be found:
\begin{equation}
\begin{array}{l}
\displaystyle Z = \frac{{{X_0}^2R}}{{{R^2} + {{({X_0} + {X_1})}^2}}} + j\frac{{{X_0}({R^2} + {X_0}{X_1} + {X_1}^2)}}{{{R^2} + {{({X_0} + {X_1})}^2}}}
\end{array}
\label{eq:3}
\end{equation}
where ${X_0} = - \frac{1}{{{C_0}\omega}}$ and ${X_1} = L\omega - \frac{1}{{C\omega}}$ represent the imaginary impedance of the static capacitor and the motional branch of the TPaD, respectively.

Since the TPaDs are excited at resonance ($\omega = \frac{1}{{\sqrt {LC} }}$), the imaginary impedance of the motional branch (${X_1}$) is equal to zero. The impedance of the TPaD system can be simplified as
\begin{equation}
\begin{array}{l}
\displaystyle Z = \frac{{{X_0}^2R}}{{{R^2} + {X_0}^2}} + j\frac{{{X_0}{R^2}}}{{{R^2} + {X_0}^2}}
\end{array}
\label{eq:1}
\end{equation}
Since the shunt resistor ($R_0$) was in series with the piezoelectric actuator, the voltage across the motional branch ($U_g$) can be expressed in terms of the input voltage, $U_i$, as follows:
\begin{equation}
\begin{array}{l}
\displaystyle {U_g} = \frac{{{U_i}}}{{\left| Z \right| + {R_0}}}\left| Z \right| = \frac{{{U_i}}}{{\frac{{{X_0}R}}{{\sqrt {{R^2} + {X_0}^2} }} + {R_0}}}\frac{{{X_0}R}}{{\sqrt {{R^2} + {X_0}^2} }}
\end{array}
\label{eq:5}
\end{equation}
Thus, the real power consumption ($\Delta P$) in the TPaDs system is equal to:
\begin{equation}
\begin{array}{l}
\displaystyle \Delta P = \frac{{{U_g}^2}}{R} = \frac{{U_i^2}}{{R{{(1 + {R_0}\sqrt {\frac{1}{{X_0^2}} + \frac{1}{{{R^2}}}} )}^2}}}
\end{array}
\label{eq:6}
\end{equation}
What is notable in this expression is that the real power consumption is expected to decrease with increasing motional resistance $R$.

\subsection{Model Results with an Artificial Finger} \label{finger_model}
\begin{figure}[htb!]
\centering
\includegraphics[width = 0.45\textwidth]{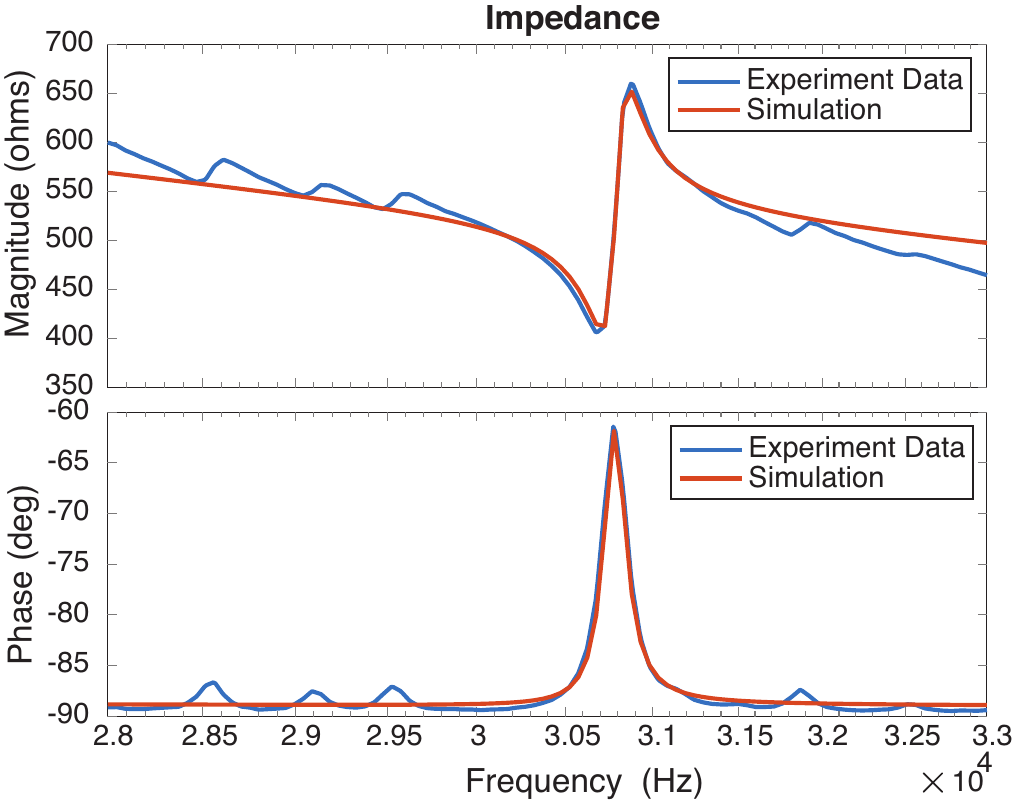}
\caption{Electrical impedance of the SLG\_0.56 without the contact of the artificial finger.}
\label{impedanceMatching}
\end{figure}

Since all the TPaDs had a similar trend, Figs. \ref{impedanceMatching}, \ref{spring:subfig}, and \ref{physical:subfig} show the performance of only the SLG\_0.56.

The electrical impedance of the SLG\_0.56 was measured by a dynamic signal analyzer (the blue curve in Fig. \ref{impedanceMatching}). It was then fitted with Eq. \ref{eq:2}. The results are shown as the red curve in Fig. \ref{impedanceMatching}. The equivalent capacitance ($C$), inductance ($L$), and resistance ($R$) of the SLG\_0.56 are shown in Fig. \ref{physical:subfig}.

When the artificial finger gradually presses on the TPaD, the equivalent capacitance decreases, but the equivalent inductance and resistance increase. The changes of the equivalent resistance are much more than that of the equivalent capacitance or inductance. This means that the damping of the finger-TPaD system increases significantly with increasing contact force. One reasonable hypothesis is that the involved damping of the artificial finger increases along with contact force. To test this hypothesis, an experiment was performed in which the artificial finger was replaced with a spring.

\subsection{Model Results with a Spring}
\begin{figure}[htb!]
  \centering
  \subfigure[]{ 
    \label{spring:subfig:a} 
    \includegraphics[width=0.45\textwidth]{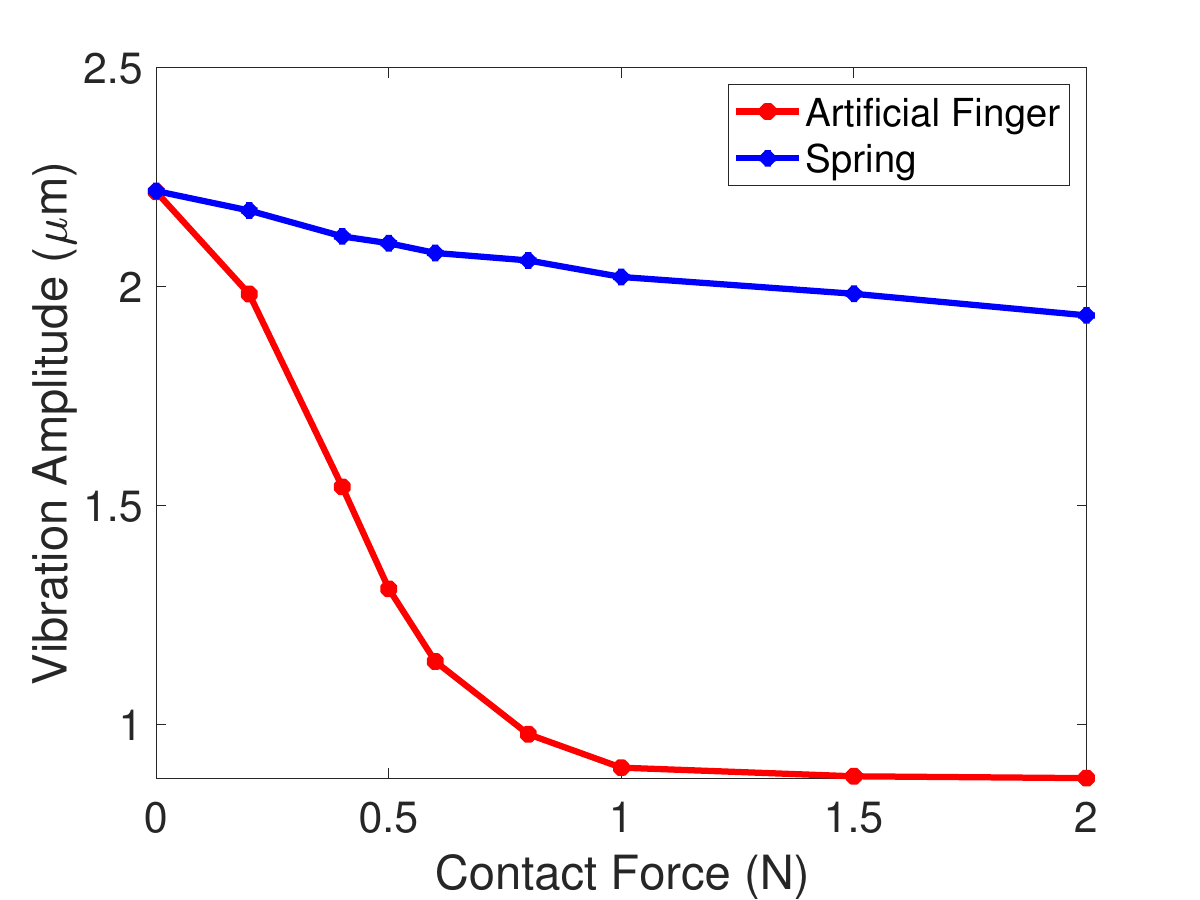}} 
  \hspace{0in} 
  \subfigure[]{ 
    \label{spring:subfig:b} 
    \includegraphics[width=0.45\textwidth]{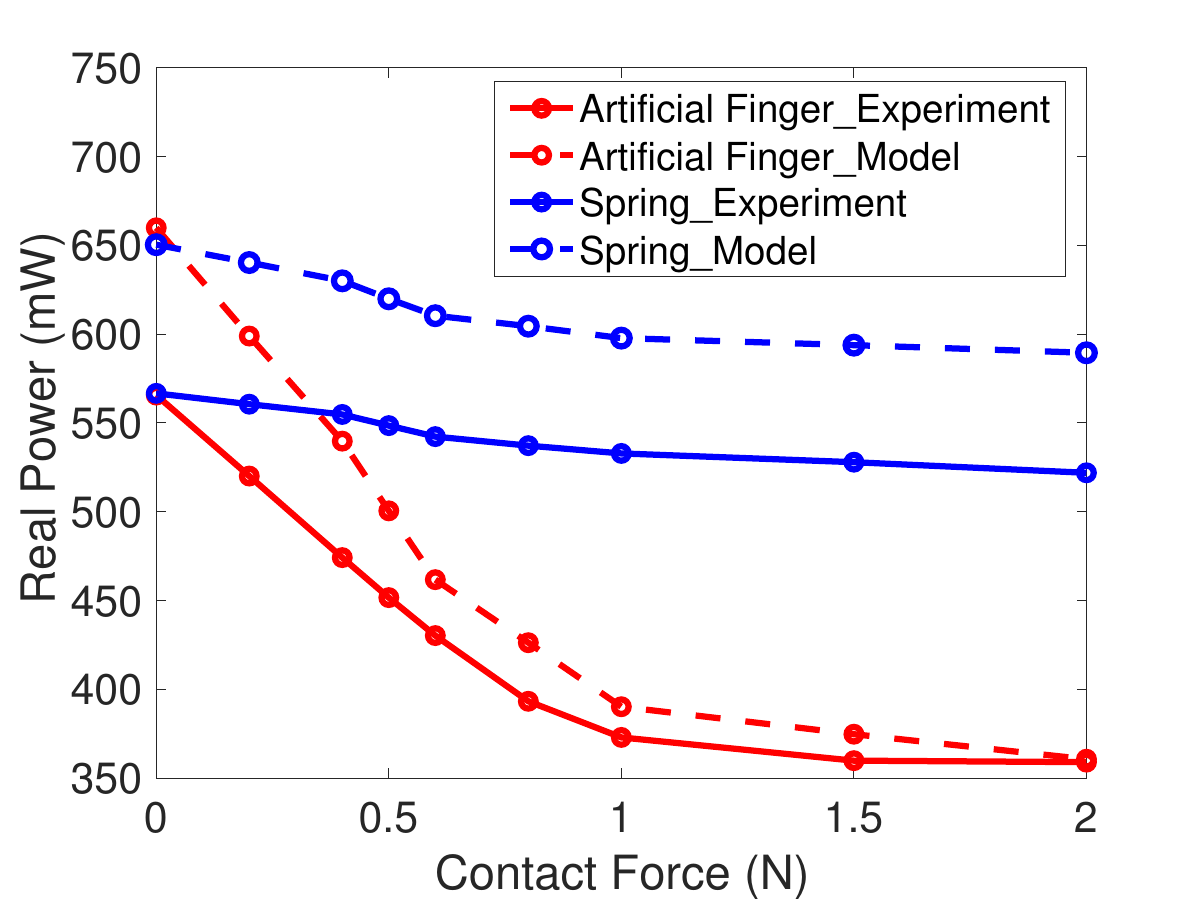}} 
  \caption{Comparisons between the SLG\_0.56 with an artificial finger and that with a spring on the vibration amplitude and the real power: (a) vibration amplitude and (b) real power.}
  \label{spring:subfig} 
\end{figure}

\begin{figure}[htb!]
  \centering
  \subfigure[]{ 
    \label{physical:subfig:a} 
    \includegraphics[width=0.45\textwidth]{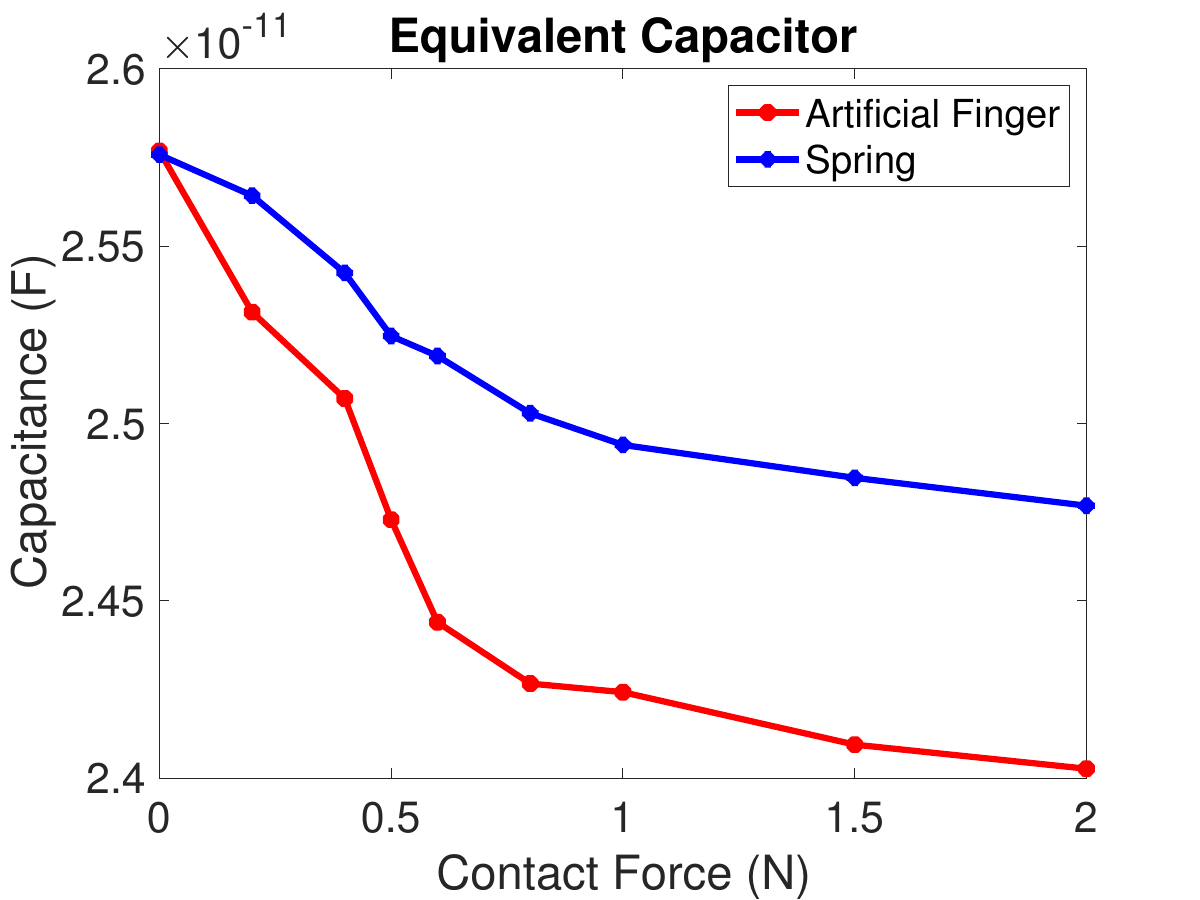}} 
  \hspace{0in} 
  \subfigure[]{ 
    \label{physical:subfig:b} 
    \includegraphics[width=0.45\textwidth]{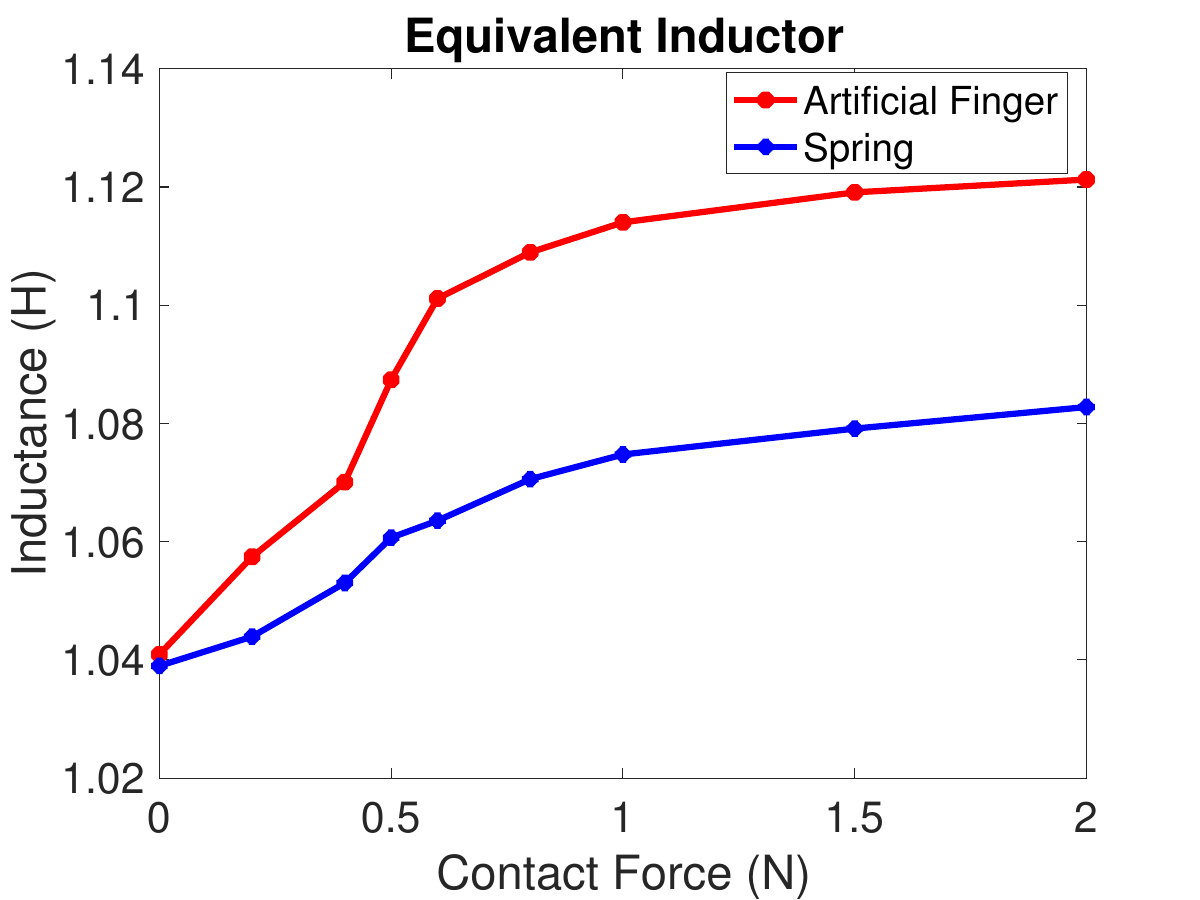}} 
  \hspace{0in} 
  \subfigure[]{ 
    \label{physical:subfig:c} 
    \includegraphics[width=0.45\textwidth]{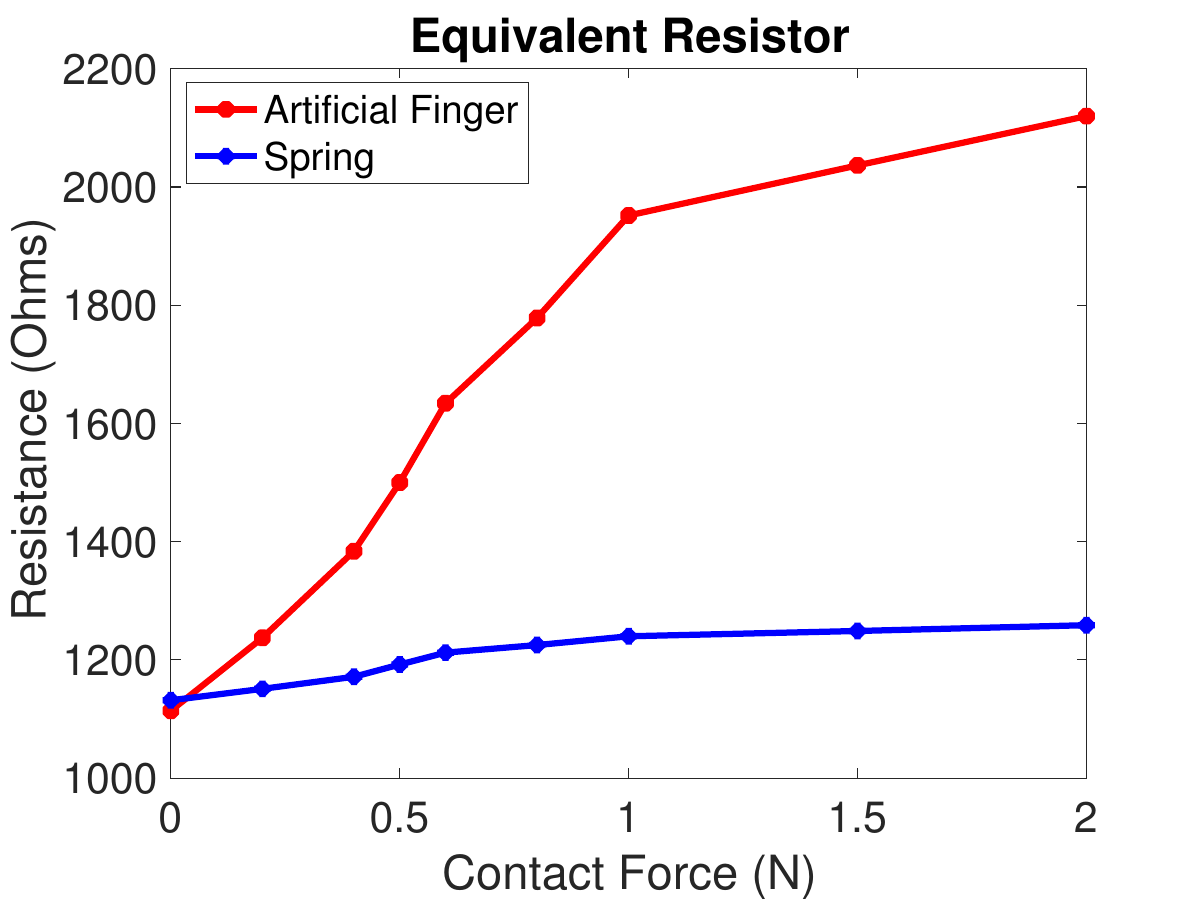}} 
  \caption{Comparison between physical changes of the SLG\_0.56 with an artificial finger and that with a spring: (a) Equivalent capacitor, (b) Equivalent inductor, and (c) Equivalent resistor.}
  \label{physical:subfig} 
\end{figure}

In Fig. \ref{spring:subfig} we see that, as the contact force increases, the revised system exhibits decreases in the vibration amplitude and the real power, just like the artificial finger, but that these decreases are much less pronounced. The equivalent $L$, $C$, and $R$ of the SLG\_0.56 with the spring were extracted by using the same fitting process as Experiment 1 (see in Fig. 9). Compared with the artificial finger, the equivalent $L$, $C$, and $R$ of the SLG\_0.56 with the spring also change more slowly (Fig. \ref{physical:subfig}). It is especially notable that the equivalent resistance of the system (1250 $\Omega $) is almost a half of that with the artificial finger (2150 $\Omega $) when the contact force is 2 N.

Eq. \ref{eq:6} and the results in Fig. \ref{physical:subfig} may be used to estimate the real power consumption for SLG\_0.56 when the artificial finger or the spring contacts the TPaD (Fig. \ref{spring:subfig:b}). As the contact force increases, the vibration amplitude decreases, consistent with the assumption that the fingertip stays in contact with the TPaD. This enables the model predictions to show good agreement with the experimental results.

\section{Discussion}
\subsection{Effects of Glass Mechanical Properties and Thickness}
Fig. \ref{materials:subfig} shows that there is a negative relationship between the glass thickness and TPaD performance (similar observations were found in \cite{giraud2010power,wang2019design}). Figs. \ref{thickness:subfig:a} and \ref{thickness:subfig:c}) indicate that a lower density and a lower Young's modulus will also modestly improve performance (see also Table \ref{Eight types of glass}). Here we consider whether the amplification number ($n$) from the piezoelectric actuator to the plate, defined in Wiertlewski et al. \cite{wiertlewski2015power}, might help explain these results. The amplification number is defined such that, as it decreases, the reflection of the plate's impedance to the actuator decreases, thereby increasing the real power consumption under a constant voltage source. 

Since Wiertlewski et al. assumed that the the actuator was made of the same material as the plate \cite{wiertlewski2015power}, we need to generalize their model in order to analyze the relation between the amplification number and the mechanical properties of the plate, including its density (${\rho_p}$), Young's modulus (${E_p}$), and thickness (${h_p}$). In order to be consistent with the descriptions in \cite{wiertlewski2015power}, we keep the same symbols for the variables. 

\begin{figure}[htb!]
\centering
\includegraphics[width = 0.45\textwidth]{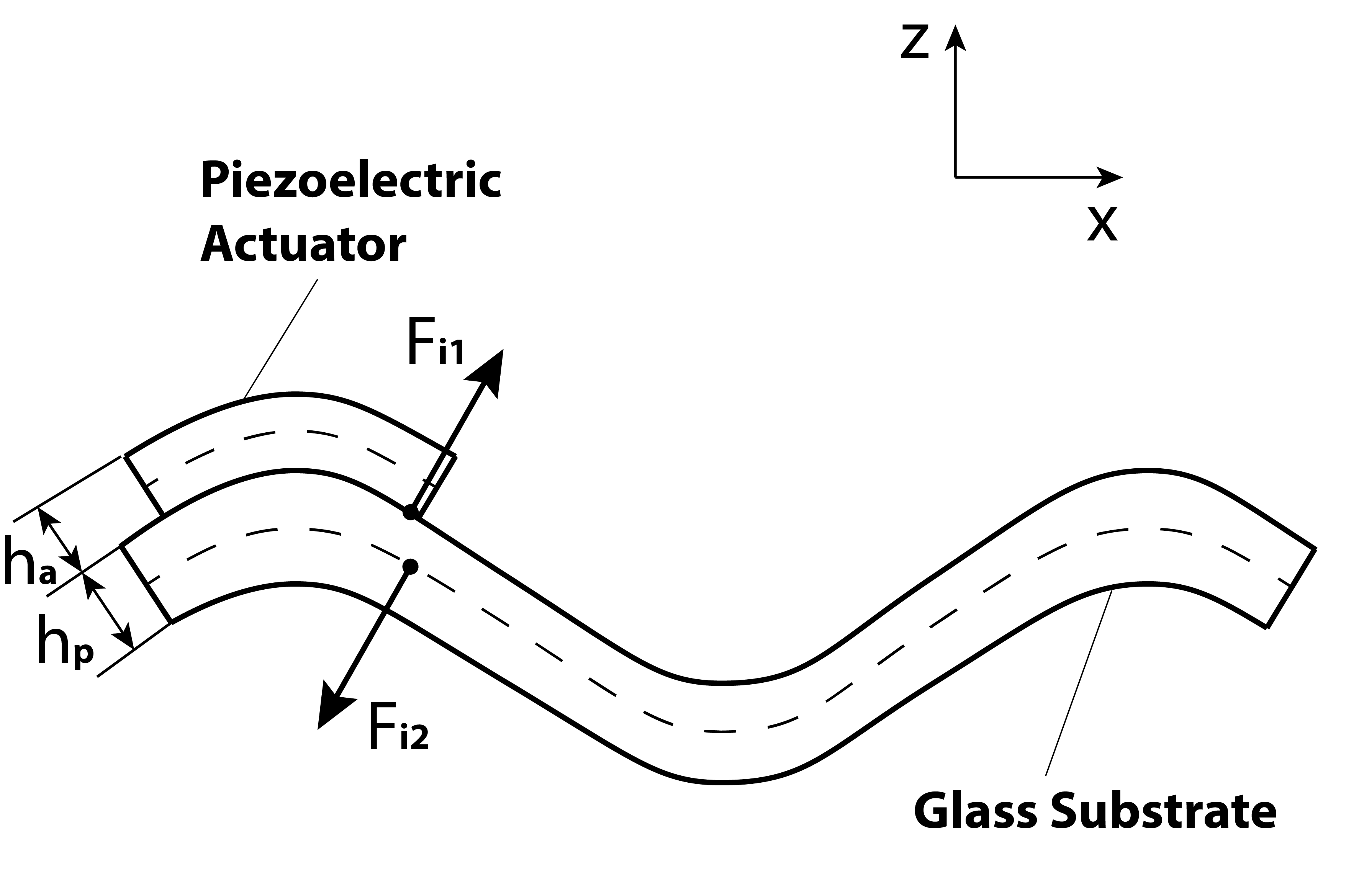}
\caption{Beam bending model of a TPaD.}
\label{beamBending}
\end{figure}

Based on Newton's third law at the junction of the actuator and the plate, the shear forces of the sandwich (the actuator and the plate) and the plate have the same magnitude but opposite directions (shown in Fig. \ref{beamBending}). 
\begin{dmath}
	\left| {{f_{i1}}} \right| = \left| {{f_{i2}}} \right|
\label{eq:12_1}
\end{dmath}
According to sandwich theory \cite{plantema1966sandwich} and Euler-Bernoulli theory, the shear forces $f_{i1}$ and $f_{i2}$ may be expressed as:
\begin{dmath}
	\left| {{f_{i1}}} \right| = D_1\frac{{{d^3}{U_a}}}{{d{x^3}}}
\label{eq:12_2}
\end{dmath}

\begin{dmath}
	\left| {{f_{i2}}} \right| = D_2\frac{{{d^3}{U_p}}}{{d{x^3}}}
\label{eq:12_3}
\end{dmath}
where the junction is assumed to be at the nodal point ($x = 0$). ${U_a}$ and ${U_p}$ are the deflections of the actuator and the plate, respectively. Thus:
\begin{dmath}
	\evalat[\Bigg]{D_1\frac{{{d^3}{U_a}}}{{d{x^3}}}}{x=0} = \evalat[\Bigg]{D_2\frac{{{d^3}{U_p}}}{{d{x^3}}}}{x=0}
\label{eq:12}
\end{dmath}
The flexural stiffness ($D_1$) of the sandwich beam is expressed as
\begin{dmath}
	D_1 = \int {\int {{z^2}E(y,z)dzdy} } = {E_p}{l_w}\frac{{h_p^3}}{3} + {E_a}{l_w}(\frac{{h_a^3}}{3} + {h_p}h_a^2 + h_p^2{h_a})
\label{eq:13}
\end{dmath}
The flexural stiffness ($D_2$) of the beam with an uniform Young's modulus is expressed as
\begin{equation}
\begin{array}{l}
\displaystyle D_2 = \int {\int {{z^2}{E_p}dzdy} } = {E_p}{I_p} = \frac{{{E_p}{l_w}h_p^3}}{{12}}
\end{array}
\label{eq:13_2}
\end{equation}
Where ${E_a}$, ${h_a}$, ${E_p}$, and ${h_p}$ represent the Young's modulus and the thickness of the actuator and the plate, respectively. $I_p$ is the second moment of area of the beam’s cross-section. ${l_w}$ is the width of the beam in y axis. Assuming ${U_a} = {u_a}\sin ({\beta _a}x)$ and ${U_p} = {u_p}\sin ({\beta _p}x)$, where ${\beta _a}$ and ${\beta _p}$ are the wavenumber of the actuator and the plate, Eq. \ref{eq:12} is simplified as
\begin{dmath}
	D_1{u_a}\beta _a^3 = \frac{{{E_p}{l_w}h_p^3}}{{12}}{u_p}\beta _p^3
\label{eq:14}
\end{dmath}
Since the amplification number is defined as the ratio of the plate amplitude to the actuator amplitude, it is represented as
\begin{equation}
\begin{array}{l}
\displaystyle n = \frac{{{u_p}}}{{{u_a}}} = \frac{{12D_1}}{{{E_p}h_p^3{l_w}}}\frac{{\beta _a^3}}{{\beta _p^3}}
\end{array}
\label{eq:15}
\end{equation}
Based on Euler–Bernoulli beam theory, the wavenumber ($\beta$) is equal to
\begin{equation}
\begin{array}{l}
\displaystyle \beta = \frac{{2\pi }}{\lambda } = \sqrt[4]{{\frac{{\mu {\omega ^2}}}{{D}}}}
\end{array}
\label{eq:16}
\end{equation}
where $\lambda$, $\mu$, and $\omega$ are the wavelength of oscillation, the mass per unit length, and the angular frequency of oscillation, respectively. Thus, the wavenumbers of the actuator ($\beta _a$) and plate ($\beta _p$) may be represented as
\begin{equation}
\begin{array}{l}
\displaystyle {\beta _a} = \sqrt[4]{{\frac{{\mu_1 {\omega ^2}}}{{D_1}}}} = \sqrt[4]{{\frac{{{\omega ^2}{l_w}({\rho _a}{h_a} + {\rho _p}{h_p})}}{D_1}}}
\end{array}
\label{eq:17}
\end{equation}

\begin{equation}
\begin{array}{l}
\displaystyle {\beta _p} = \sqrt[4]{{\frac{{\mu_2 {\omega ^2}}}{{D_2}}}} = \sqrt[4]{{\frac{{\mu_2 {\omega ^2}}}{{{E_p}{I_p}}}}} = \sqrt[4]{{\frac{{12{\omega ^2}{\rho _p}}}{{{E_p}h_p^2}}}}
\end{array}
\label{eq:18}
\end{equation}
Eq. \ref{eq:15} may now be simplified to
\begin{dmath}
	n = 12 * {[\frac{1}{{12}}{(\frac{{D_1'}}{{{E_p}}})^{\frac{1}{3}}}(\frac{{{\rho _a}{h_a}}}{{h_p^2{\rho _p}}} + \frac{1}{{{h_p}}})]^{\frac{3}{4}}}
\label{eq:19}
\end{dmath}
Where $D_1'$ is $D_1$ divided by $l_w$.
\begin{dmath}
	D_1' = {E_p}\frac{{h_p^3}}{3} + {E_a}(\frac{{h_a^3}}{3} + {h_p}h_a^2 + h_p^2{h_a})
\label{eq:20}
\end{dmath}

\begin{figure}[htb!]
  \centering
  \subfigure[]{ 
    \label{amplification:subfig:a} 
    \includegraphics[width=0.45\textwidth]{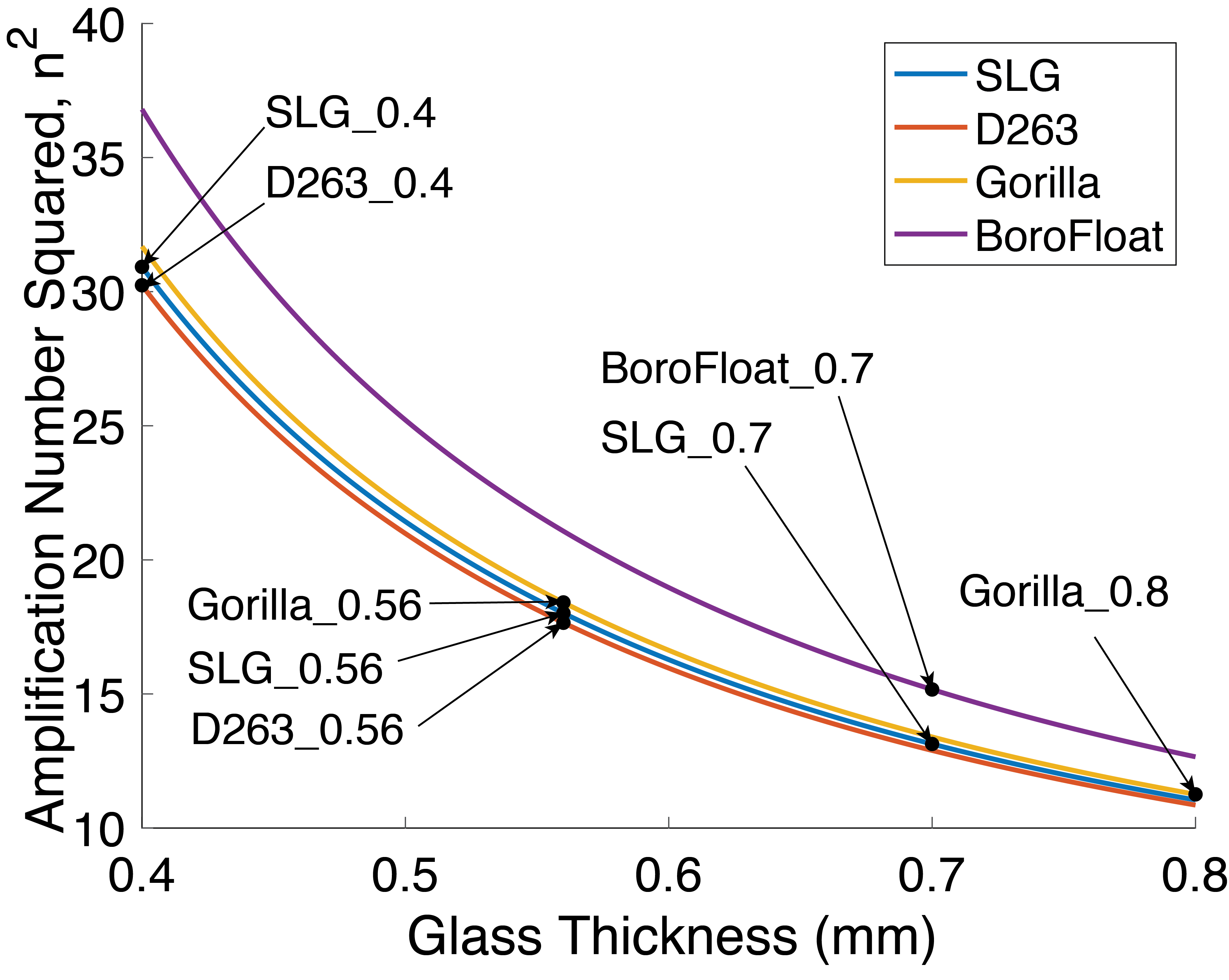}} 
  \hspace{0in} 
  \subfigure[]{ 
    \label{amplification:subfig:b} 
    \includegraphics[width=0.45\textwidth]{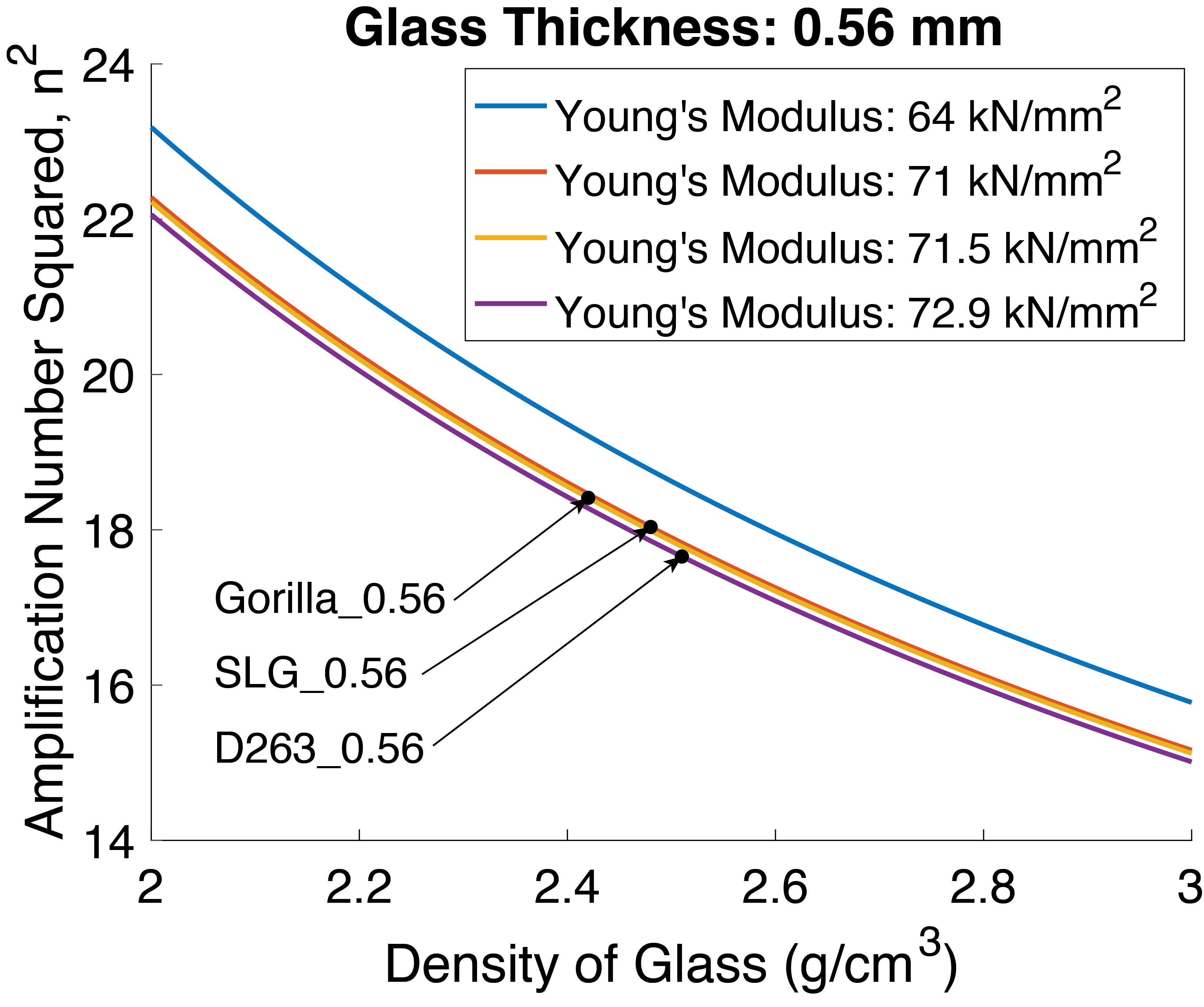}} 
  \hspace{0in} 
  \subfigure[]{ 
    \label{amplification:subfig:c} 
    \includegraphics[width=0.45\textwidth]{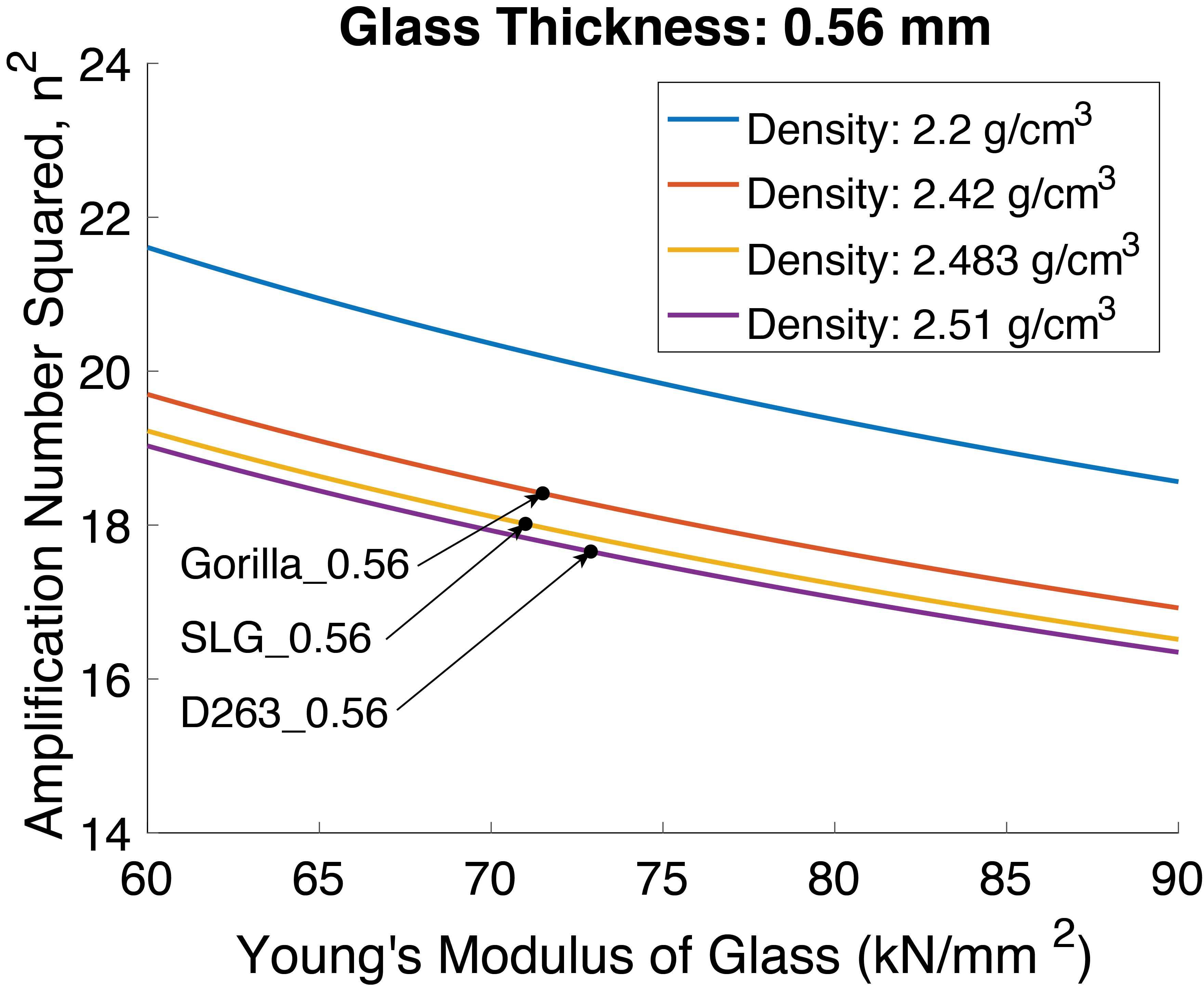}} 
  \caption{Relation between the square of the amplification number ($n^2$) and the thickness ($h_p$), density ($\rho _p$), Young's modulus ($E _p$) of glass: (a) Thickness ($h_p$), (b) Density ($\rho _p$), and (c) Young's modulus ($E _p$).}
  \label{amplification:subfig} 
\end{figure}

A hard piezoelectric actuator (SMPL60W5T03R112, Steminc and Martins Inc, Miami, FL, USA) was used for each TPaD. Its thickness ($h_a$), density ($\rho_a$), and Young's Modulus ($E_a$) are 0.3 mm, 7.9 $g/cm^3$, and 84 $kN/mm^2$, respectively. Using these properties as well as those of the various TPaDs, the relation between the square of the amplification number ($n^2$) and the thickness ($h_p$), density ($\rho _p$), and Young's modulus ($E _p$) of glass are shown in Fig. \ref{amplification:subfig}.

Assuming that the reflected impedance of the glass ($n^2 Z_{glass}$) is much greater than the actuator impedance ($Z_{actuator}$), and assuming that the glass impedances ($Z_{glass}$) of all the TPaDs are similar, the real power consumption $\Delta {P}$ of any two TPaDs under a constant voltage supply may be compared as follows:
\begin{equation}
	\frac{{\Delta {P_2}}}{{\Delta {P_1}}} = \frac{{n_1^2{Z_{glass}}}}{{n_2^2{Z_{glass}}}} = \frac{{n_1^2}}{{n_2^2}}
\label{eq:21}
\end{equation}
where $\Delta {P_1}$, ${n_1}$, $\Delta {P_2}$, and ${n_2}$ are the real power consumption and the amplification number of one type of TPaD and another type of TPaD, respectively. Thus, as the amplification number decreases, the real power consumption increases (assuming a constant voltage source).

\begin{figure}[htb!]
\centering
\includegraphics[width = 0.45\textwidth]{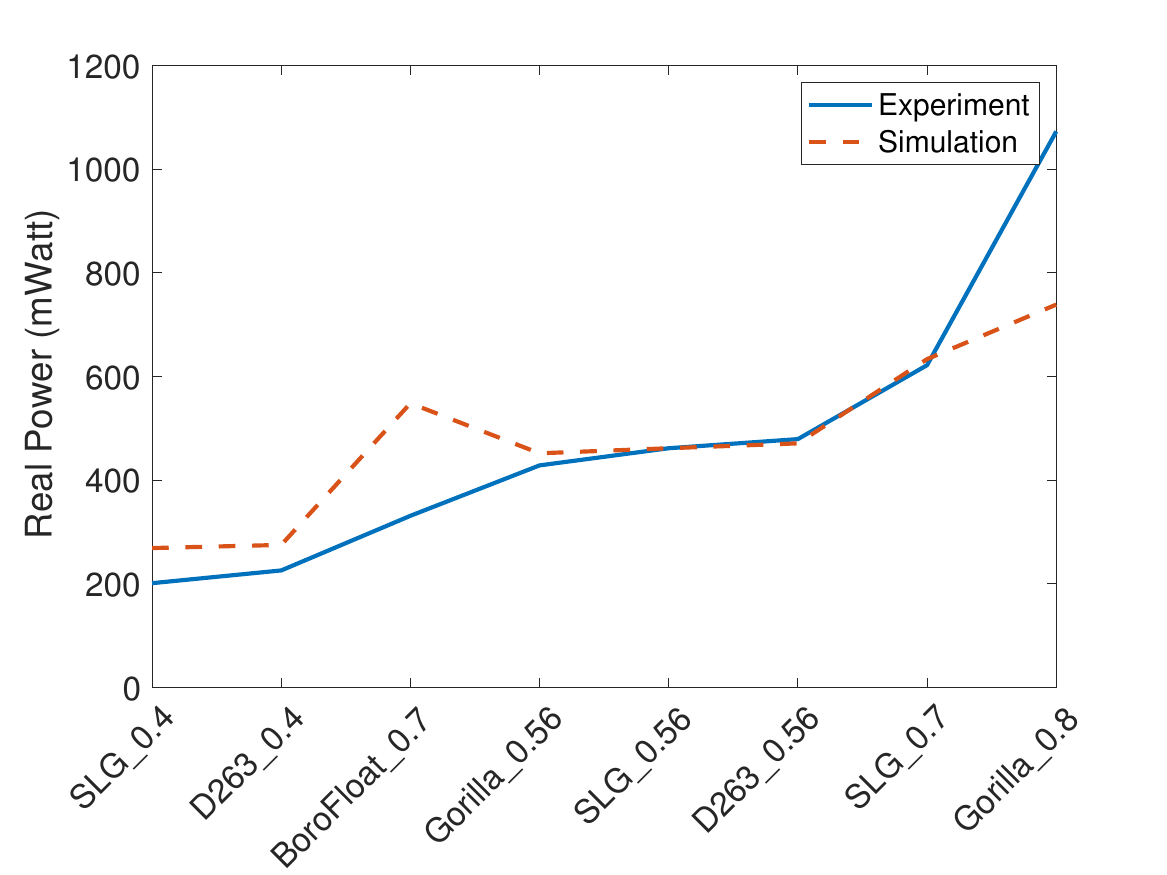}
\caption{Comparison of real power consumption between experiment results in Fig. \ref{all:subfig:b} and simulations based on Eq. \ref{eq:21}.}
\label{realPower_simulation}
\end{figure}

Eq. \ref{eq:21} may be used to compare the expected power consumption of the various TPaDs against that of one reference, which is taken to be SLG\_0.56. This is shown in Fig. \ref{realPower_simulation}) and provides generally quite good agreement with experimental results. Thus, the effects of thickness, density, and Young's modulus on the amplification number could be reasons why TPaDs with thick, dense, or stiff glass exhibit lower power efficiency in Figs. \ref{materials:subfig} and \ref{thickness:subfig}.

\subsection{Summary: Changes in the Real Power} \label{Changes_in_the_Real_Power}
Several factors have been shown to affect power consumption. Notably, an increase in damping will decrease real power consumption under constant voltage excitation (Eq. \ref{eq:6}). One consequence of this is that increasing contact force with a finger will decrease power (presumably because damping grows as the contact area increases), while increasing contact force with a spring will have minimal effect. Additionally, an increase in reflected impedance due to thinner, less dense or less stiff glass, will reduce power consumption.

\section{Conclusions}
This study has contributed to the design of surface haptic devices based on transverse ultrasonic vibrations by elucidating the manner in which material and geometric properties of a glass plate will affect friction reduction as well as real power consumption. Three leading ultrasonic friction reduction models were reviewed and a rationale was provided for adopting the model of Vezzoli, et al. and Sednaoui, et al. \cite{vezzoli2017friction,sednaoui2017friction}. This model and a set of very carefully constructed TPaDs were used to investigate performance. The experimental results, an electromechanical model, and a mechanics-based estimate of reflected impedance, collectively shed light on the underlying factors affecting vibration velocity magnitude and power consumption. These results allow for design optimization strategies, such as using glass that is thin with a low Young’s modulus and low density. Future work can further address design optimization including actuator parameters that affect the amplification number. Additionally, it may be interesting to apply these results to better understand why TPaD performance can vary considerably from person to person.

\ifCLASSOPTIONcompsoc
  \section*{Acknowledgments}
\else
  \section*{Acknowledgment}
\fi

This material is based upon work supported by the National Science Foundation grant number IIS-1518602.

\ifCLASSOPTIONcaptionsoff
  \newpage
\fi

\bibliographystyle{IEEEtran}
\bibliography{reference}

\begin{thebibliography}{10}
\providecommand{\url}[1]{#1}
\csname url@samestyle\endcsname
\providecommand{\newblock}{\relax}
\providecommand{\bibinfo}[2]{#2}
\providecommand{\BIBentrySTDinterwordspacing}{\spaceskip=0pt\relax}
\providecommand{\BIBentryALTinterwordstretchfactor}{4}
\providecommand{\BIBentryALTinterwordspacing}{\spaceskip=\fontdimen2\font plus
\BIBentryALTinterwordstretchfactor\fontdimen3\font minus
  \fontdimen4\font\relax}
\providecommand{\BIBforeignlanguage}[2]{{%
\expandafter\ifx\csname l@#1\endcsname\relax
\typeout{** WARNING: IEEEtran.bst: No hyphenation pattern has been}%
\typeout{** loaded for the language `#1'. Using the pattern for}%
\typeout{** the default language instead.}%
\else
\language=\csname l@#1\endcsname
\fi
#2}}
\providecommand{\BIBdecl}{\relax}
\BIBdecl

\bibitem{bau2010teslatouch}
O.~Bau, I.~Poupyrev, A.~Israr, and C.~Harrison, ``Teslatouch: electrovibration
  for touch surfaces,'' in \emph{Proceedings of the 23nd annual ACM symposium
  on User interface software and technology}.\hskip 1em plus 0.5em minus
  0.4em\relax ACM, 2010, pp. 283--292.

\bibitem{linjama2009sense}
J.~Linjama and V.~M{\"a}kinen, ``E-sense screen: Novel haptic display with
  capacitive electrosensory interface,'' \emph{Demo paper Submitted for HAID},
  vol.~9, pp. 10--11, 2009.

\bibitem{shultz2015surface}
C.~D. Shultz, M.~A. Peshkin, and J.~E. Colgate, ``Surface haptics via
  electroadhesion: expanding electrovibration with johnsen and rahbek,'' in
  \emph{World Haptics Conference (WHC), 2015 IEEE}.\hskip 1em plus 0.5em minus
  0.4em\relax IEEE, 2015, pp. 57--62.

\bibitem{meyer2014dynamics}
D.~J. Meyer, M.~Wiertlewski, M.~A. Peshkin, and J.~E. Colgate, ``Dynamics of
  ultrasonic and electrostatic friction modulation for rendering texture on
  haptic surfaces.'' in \emph{HAPTICS}, 2014, pp. 63--67.

\bibitem{salbu1964compressible}
E.~Salbu, ``Compressible squeeze films and squeeze bearings,'' \emph{Journal of
  Basic Engineering}, vol.~86, no.~2, pp. 355--364, 1964.

\bibitem{winfield2008variable}
L.~Winfield and J.~Colgate, ``Variable friction haptic displays,'' \emph{Haptic
  rendering: Foundations, algorithms and applications, MC Lin and M. Otaduy,
  Eds. AK Peters, Ltd}, 2008.

\bibitem{wiertlewski2016partial}
M.~Wiertlewski, R.~F. Friesen, and J.~E. Colgate, ``Partial squeeze film
  levitation modulates fingertip friction,'' \emph{Proceedings of the National
  Academy of Sciences}, vol. 113, no.~33, pp. 9210--9215, 2016.

\bibitem{watanabe1995method}
T.~Watanabe and S.~Fukui, ``A method for controlling tactile sensation of
  surface roughness using ultrasonic vibration,'' in \emph{Robotics and
  Automation, 1995. Proceedings., 1995 IEEE International Conference on},
  vol.~1.\hskip 1em plus 0.5em minus 0.4em\relax IEEE, 1995, pp. 1134--1139.

\bibitem{biet2007squeeze}
M.~Biet, F.~Giraud, and B.~Lemaire-Semail, ``Squeeze film effect for the design
  of an ultrasonic tactile plate,'' \emph{IEEE transactions on ultrasonics,
  ferroelectrics, and frequency control}, vol.~54, no.~12, pp. 2678--2688,
  2007.

\bibitem{winfield2007t}
L.~Winfield, J.~Glassmire, J.~E. Colgate, and M.~Peshkin, ``T-pad: Tactile
  pattern display through variable friction reduction,'' in \emph{Second Joint
  EuroHaptics Conference and Symposium on Haptic Interfaces for Virtual
  Environment and Teleoperator Systems (WHC'07)}.\hskip 1em plus 0.5em minus
  0.4em\relax IEEE, 2007, pp. 421--426.

\bibitem{wiertlewski2015power}
M.~Wiertlewski and J.~E. Colgate, ``Power optimization of ultrasonic
  friction-modulation tactile interfaces,'' \emph{IEEE transactions on
  haptics}, vol.~8, no.~1, pp. 43--53, 2015.

\bibitem{friesen2016role}
R.~F. Friesen, M.~Wiertlewski, and J.~E. Colgate, ``The role of damping in
  ultrasonic friction reduction,'' in \emph{2016 IEEE Haptics Symposium
  (HAPTICS)}.\hskip 1em plus 0.5em minus 0.4em\relax IEEE, 2016, pp. 167--172.

\bibitem{vezzoli2017friction}
E.~Vezzoli, Z.~Vidrih, V.~Giamundo, B.~Lemaire-Semail, F.~Giraud, T.~Rodic,
  D.~Peric, and M.~Adams, ``Friction reduction through ultrasonic vibration
  part 1: Modelling intermittent contact,'' \emph{IEEE transactions on
  haptics}, vol.~10, no.~2, pp. 196--207, 2017.

\bibitem{persson2013contact}
B.~Persson, A.~Kovalev, and S.~Gorb, ``Contact mechanics and friction on dry
  and wet human skin,'' \emph{Tribology Letters}, vol.~50, no.~1, pp. 17--30,
  2013.

\bibitem{sednaoui2017friction}
T.~Sednaoui, E.~Vezzoli, B.~Dzidek, B.~Lemaire-Semail, C.~Chappaz, and
  M.~Adams, ``Friction reduction through ultrasonic vibration part 2:
  Experimental evaluation of intermittent contact and squeeze film
  levitation,'' \emph{IEEE transactions on haptics}, vol.~10, no.~2, pp.
  208--216, 2017.

\bibitem{giraud2018evaluation}
F.~Giraud, T.~Hara, C.~Giraud-Audine, M.~Amberg, B.~Lemaire-Semail, and
  M.~Takasaki, ``Evaluation of a friction reduction based haptic surface at
  high frequency,'' in \emph{2018 IEEE Haptics Symposium (HAPTICS)}.\hskip 1em
  plus 0.5em minus 0.4em\relax IEEE, 2018, pp. 210--215.

\bibitem{xu2019ultrashiver}
H.~Xu, M.~A. Peshkin, and E.~Colgate, ``Ultrashiver: Lateral force feedback on
  a bare fingertip via ultrasonic oscillation and electroadhesion,'' \emph{IEEE
  transactions on haptics}, 2019.

\bibitem{giraud2010power}
F.~Giraud, M.~Amberg, R.~Vanbelleghem, and B.~Lemaire-Semail, ``Power
  consumption reduction of a controlled friction tactile plate,'' in
  \emph{International Conference on Human Haptic Sensing and Touch Enabled
  Computer Applications}.\hskip 1em plus 0.5em minus 0.4em\relax Springer,
  2010, pp. 44--49.

\bibitem{wang2019design}
S.~Wang, X.~Lu, and H.~Sun, ``Design and optimization of squeeze film effect
  based tactile display system,'' in \emph{2019 IEEE World Haptics Conference
  (WHC)}.\hskip 1em plus 0.5em minus 0.4em\relax IEEE, 2019, pp. 169--174.

\bibitem{plantema1966sandwich}
F.~J. Plantema, ``Sandwich construction: the bending and buckling of sandwich
  beams, plates, and shells,'' 1966.

\end{thebibliography}

\begin{IEEEbiography}[{\includegraphics[width=1in,height=1.25in,clip,keepaspectratio]{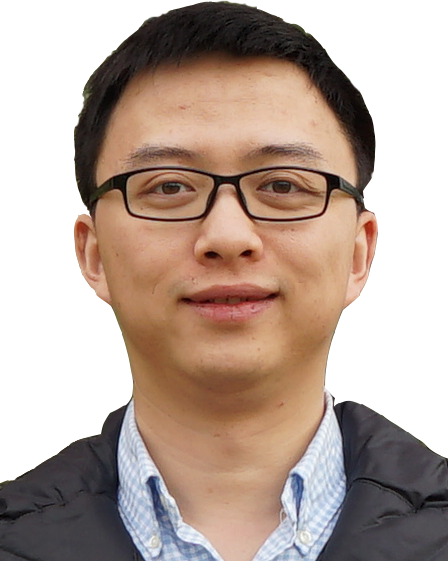}}]{Heng Xu}
is a Ph.D. candidate in the Department of Mechanical Engineering, Northwestern University, Evanston, IL, USA. His research interests include surface haptics, electrotactile stimulation, and tactile sensory feedback. He is a member of the IEEE.
\end{IEEEbiography}

\begin{IEEEbiography}[{\includegraphics[width=1in,height=1.25in,clip,keepaspectratio]{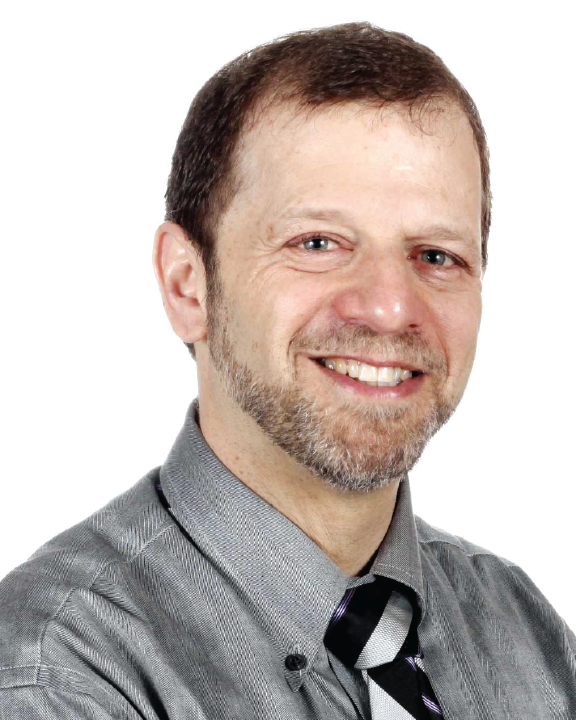}}]{Michael A. Peshkin}
is the Bette and Neison Harris professor of teaching excellence in the Department of Mechanical Engineering, Northwestern University, Evanston, Illinois. His research is in haptics, robotics, human-machine interaction, and rehabilitation robotics. He has co-founded four start-up companies: Mako Surgical, Cobotics, HDT Robotics, and Tanvas. He is a fellow of the National Academy of Inventors, and (with J. E. Colgate) an inventor of cobots. He is a senior member of the IEEE.
\end{IEEEbiography}

\begin{IEEEbiography}[{\includegraphics[width=1in,height=1.25in,clip,keepaspectratio]{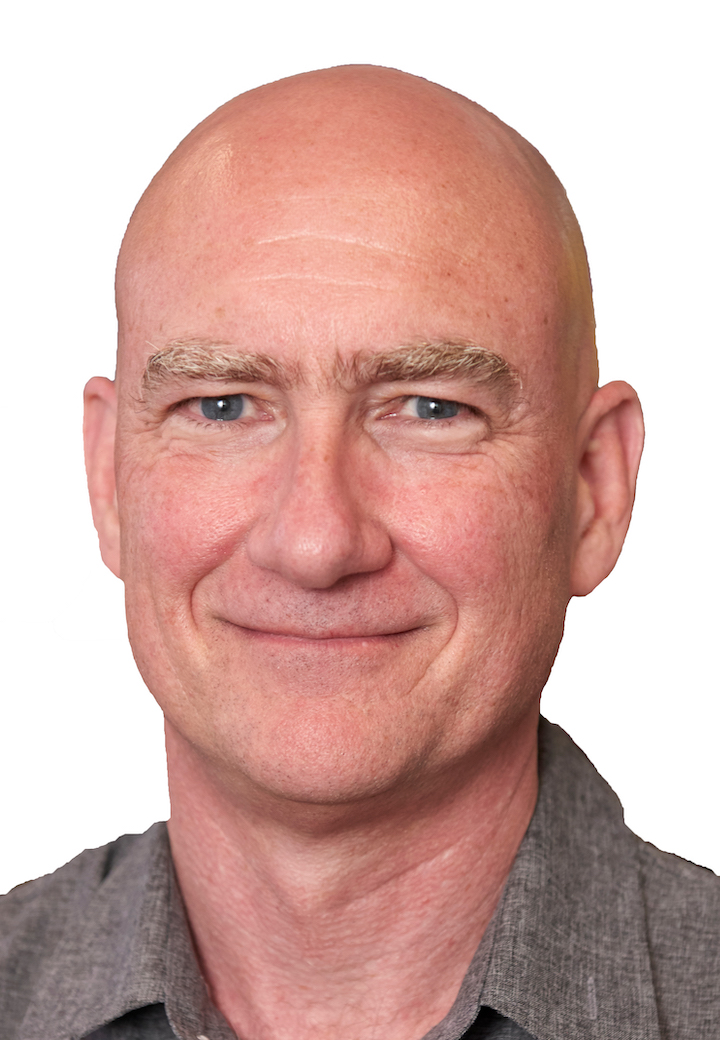}}]{J. Edward Colgate}
is the Breed University Design professor at Northwestern University. He is known for his work in haptics and human-robot collaboration. He served as an associate editor of the IEEE Transactions on Robotics and Automation, and he was the founding editor-in-chief of the IEEE Transactions on Haptics. He was one of the founding codirectors of the Segal Design Institute, Northwestern University. He has founded three start-up companies the most recent of which, Tanvas Inc., is commercializing
an innovative haptic touch screen. He is a fellow of the IEEE and the National Academy of Inventors.
\end{IEEEbiography}

\end{document}